\documentclass[twocolumn]{IEEEtran}
\usepackage{amsfonts}
\usepackage{amssymb}
\usepackage{amsmath}
\usepackage{graphicx}
\usepackage{color}
\usepackage{multirow}
\usepackage{cite}
\usepackage{stfloats}
\usepackage[font=small]{caption}
\usepackage{subcaption}
\usepackage{algorithm}
\usepackage{algorithmic}
\usepackage{array}
\usepackage{bbm}
\usepackage{epstopdf}

\graphicspath{{../}}
\input{tcilatex}

\newtheorem{rema}{Remark}

\begin{document}

\title{Enhancing Cellular Performance through Device-to-Device Distributed
MIMO\thanks{}}
\author{{Jiajia Guo, \textit{Student member, IEEE}, Wei Yu, \textit{Fellow, IEEE}, and Jinhong Yuan \textit{Fellow, IEEE}} \thanks{%
Manuscript to appear in IEEE Transactions on Communications.
J. Guo was with the School of Electrical Engineering and Telecommunication,
University of New South Wales, Sydney, NSW 2052, Australia,
and Data61, CSIRO, Marsfield, NSW 2122, Australia. She is now with
California Partners for Advanced Transportation Technology (PATH), at
the University of California, Berkeley, Berkeley CA 94720 USA (e-mail:
jiajia.guo@berkeley.edu).
W. Yu is with the University of Toronto, Toronto, ON M5S 3G4, Canada
(e-mail: weiyu@comm.utoronto.ca).
J. Yuan is with the University of New South Wales, Sydney, NSW 2052,
Australia (e-mail: j.yuan@unsw.edu.au).}}
\maketitle

\begin{abstract}
The integration of local device-to-device (D2D) communications and cellular
connections has been intensively studied to satisfy co-existing D2D and
cellular communication demand. In future cellular networks, there will be
numerous standby users possessing D2D communication capabilities in close
proximity to each other. Considering that these standby users do not
necessarily request D2D communications all the time, in this paper we propose a
hybrid D2D-cellular scheme to make use of these standby users and to improve
the rate performance for cellular users. More specifically, through D2D
links, a virtual antenna array can be formed by sharing antennas across
different terminals to realize the diversity gain of MIMO channels. This
paper considers the use of millimeter wave (mmWave) links to enable high
data rate D2D communications. We then design an orthogonal D2D multiple access protocol and
formulate the optimization problem of joint cellular and D2D resource
allocation for downlink transmissions using the proposed scheme. We obtain a
closed-form solution for D2D resource allocation, which reveals useful
insights for practical system design. Numerical results from extensive
system-level simulations demonstrate that the rate performance of cellular
users is significantly improved.
\end{abstract}


\section{Introduction}

The fifth generation (5G) wireless networks are expected to meet the booming
data traffic demand and to provide a better quality of service and user
experience. The integration of device-to-device (D2D) communications into a
traditional cellular network has been proposed as one of the promising
technologies for 5G networks \cite{DopplerM09,FengCM14}. D2D communications
in cellular networks are mainly considered for local traffic handling, which
enables direct communications between two mobile users without traversing
the Base Station (BS) or core network. Studies have shown that this hybrid
network can significantly increase network spectral efficiency (SE) and
energy efficiency (EE) and alleviate the core network congestion\cite%
{RatasukM14}.

However, we note that most current works are dedicated to increasing the SE
and EE from the resource utilization point of view, instead of improving the
cellular user performance. A typical scenario under the current D2D-cellular
framework is that two user terminals communicate with each other
directly through D2D links. But this is not the only scenario where D2D
communication is useful. In this paper, we make an observation that with a
large number of end-users in the network, there are numerous devices
possessing D2D communication capabilities, but do not have a need of
performing communications at any given time. We refer to the users who do
not request cellular connections or D2D communications as \emph{standby users%
}. Our idea here is to make use of the D2D capability of these standby users
to improve the active cellular user performance. Since we only use standby
users without D2D communication demand, our proposal can be applied on top
of current D2D framework in a dense cellular network.

\subsection{Related Works and Motivation}

\subsubsection{D2D communications}

D2D communications in cellular networks can take place in the cellular
spectrum (i.e., inband) or unlicensed spectrum (i.e., outband). The majority
of current literature proposes to use the cellular spectrum for both D2D and
cellular communications (i.e., underlay inband D2D). The arising co-channel
interference between cellular users and D2D users is a major issue.
Sophisticated resource allocation methods can be used to alleviate
interference and to improve SE and EE, but most of them have high complexity
\cite{XuJSAC13,FengTCOM13,GhoshTWC14}. The orthogonalization of D2D and
cellular communications over the cellular spectrum (i.e., overlay inband
D2D) can completely avoid such interference \cite{LiangTWC13}\cite{ZhouVTC13}%
, but at the expense of a lower SE improvement as compared to underlay
proposals.

Outband D2D genuinely enjoys interference-free transmissions for cellular
and D2D users, which largely simplifies the user management at the BS \cite%
{NgTWC16}. However, it has been widely accepted that the traditional D2D
technologies are inadequate. First, the widely known D2D technologies, such
as Bluetooth and WiFi, work at the 2.4GHz unlicensed band. This band is
rather crowded and thus the interference is uncontrollable. In addition,
both Bluetooth and WiFi require manual pairing between two devices, which
causes inconvenience in customer experience \cite{HongTWC11}.

A promising technology of the future 5G networks is millimeter wave (mmWave)
communication, providing multi-gigabits-per-second to the end users \cite%
{AndrewsJSAC14}, making them potentially applicable to D2D communications.
More importantly, the interference problem would be largely alleviated due
to the highly directional antennas and large propagation loss in mmWave
communications. In several proposed mmWave D2D communications, the pairing
of D2D devices can be handled by BSs thus providing better user
experience. For these reasons, this paper proposes to use mmWave D2D
communications to complement current microwave cellular networks.

\subsubsection{Cooperative communications}

Cooperative communications have been intensively studied in literatures and
have been regarded as a focal technology in the cellular networks today \cite%
{LinJSAC97}\cite{BinAccess14}. In general, D2D communications can be
utilized for cooperative communications, e.g., packet forwarding and
relaying, and their impact is expected to be remarkable \cite{FitzekWPC09-PP}%
. A major application of D2D relaying in cellular networks is multi-hop,
which has been recognized to be capable of reducing transmission power and
increasing network capacity. For example, when the D2D pairs are far away
from each other, the direct link between the users is not good enough for
communication and other devices can then relay signals between the D2D pairs
\cite{NishiyamaM14}. In \cite{WangTWC15}\cite{ZhangTWC09}, multihop has also
been proposed to enhance the cellular transmissions, where the user with a
strong channel can forward the received message from the BS to a weak user
via D2D links. This kind of relaying introduces so-called \textquotedblleft
hop gain\textquotedblright , which can be seen as a power gain.

Another way of relaying is to exploit the diversity gain through the
distributed multiple-input and multiple-output (MIMO) technology. In \cite%
{MThesis}, the concept of a virtual antenna array was proposed, where mobile
users are clustered to form a virtual antenna array which emulates a MIMO
device via D2D communications. A similar concept called \textquotedblleft
distributed antenna systems\textquotedblright\ has also gained attentions
and has been shown to be able to cover the dead spots in wireless networks,
extend service coverage, improve spectral efficiency and mitigate
interference \cite{HeathM13}\cite{ZhuM12}. Specifically, such a system
architecture can realize the potential diversity gain by sharing antennas
across the different terminals to form a virtual MIMO system \cite%
{JiangTVT11}. The authors of \cite{NgTWC16} proposed a shared user
equipment-side distributed antenna system, which utilizes mmWave D2D links
to enable a spatial multiplexing gain for single-antenna end users to
improve the energy efficiency of outdoor-to-indoor communications. \cite{LeeTCom16} studied cooperations between active cellular devices, and thus focused on how to pair these devices such that the average throughput is maximized. Yet,
studies on this direction remains limited.

Most current studies tend to treat the D2D links as ideal high-rate
connections and focus on the allocation of cellular resources \cite{NgTWC16}%
\cite{MThesis}. We note that D2D connections are over wireless medium and
D2D channel conditions may widely vary from one link to another due to
fading and path loss, resulting in different D2D link capacities. In
addition, different standby users may make different contributions to the
rate of the destination users due to independent cellular channel
realizations. Based on the above considerations, the following questions
arise: i) which standby users should be enabled, and ii) how many D2D
resources should be allocated to different standby users. These questions
have been dealt with in \cite{ZhouVTC13}. However, their proposed algorithm
is based on a multi-hop D2D-cellular scheme.

For a MIMO relay scheme, the joint optimization of the MIMO transmission and
relaying strategies is a difficult task, because the information theoretical
capacity of the relay channel is still an open problem \cite{KramerIT05}.
The authors of \cite{ArvinICC15} consider the scenario of a single cellular
user with a single multi-antenna relay and then address the problem of
jointly optimizing the transmit covariance and quantization noise covariance
matrix. Their results confirm that optimized relaying can lead to
significant throughput improvement. In order to simplify cooperative
transmissions and reduce complexity at each relay (standby user), our scheme
performs independent individual quantization at multiple single-antenna
relays, which is different from the joint quantization at a single
multi-antenna relay in \cite{ArvinICC15}. Besides, individual
relay-destination capacity constraints are considered instead of a sum
relay-destination capacity constraint. In addition, we use the sum rate of
multi-cell cellular users as our performance merit instead of a single user
rate. These considerations impose new challenges in the joint optimizing
MIMO transmission and relaying.

\subsubsection{Cellular networks}

Classic communication theory has revealed great benefits of equipping
multiple antennas on users in wireless channels \cite{SpencerTSP04,LiIT12,LiuIT15}.
Especially, the multiple antennas are able to enhance the capability of
encountering inter-cell interferences, which accounts for the main cause of
rate degradation in dense cellular networks. However, with the size limit of
user equipment, it is difficult to equip the end users with many antennas in
the current micro-wave cellular system and for this reason, comparative
studies on the benefit of user side MIMO are limited. For example, in a
multi-cell interfering broadcast channel, with multiple antennas at
end-users, how many users should be scheduled as a function of transmit
antennas at BS? How much performance gain can be obtained in practical
systems? These questions have not yet been answered to the best of our
knowledge. With the virtual antenna array formed via D2D links, multiple
antennas at end-users in micro-wave cellular system become a real
possibility. This paper examines these questions and quantifies the benefits
of the proposed scheme through system-level simulations.

\subsection{Our Contributions}

While the majority of current D2D-cellular works focus on the co-existing of
D2D and cellular communications demand, this paper instead aims to employ
D2D communications to improve the data rate and enhance the user experience
of cellular users. Toward this end, this paper focuses on the multi-cell
interfering broadcast channel. We summarize our contributions as follows.

\begin{itemize}
\item We propose a hybrid D2D-cellular scheme applying the distributed MIMO
technology to assist the cellular communications in 5G networks. With a
user-clustering strategy, the active cellular user and its nearby standby
users form a \emph{cluster}. A MIMO channel is then emulated through D2D
links from the BS to the cluster, providing a \emph{diversity gain} for
single-antenna end users. This is different from the existing multi-hop approach, which essentially provides a \emph{power gain}. The virtual MIMO channel enhances the capability of inter-cell
interference mitigation at end users, and thus improves the rate
performance, especially for cell-edge users.

\item We employ mmWave communications for D2D links to avoid interference
from cellular transmissions. We then propose an orthogonal D2D multiple
access protocol to manage D2D interferences. The proposed protocol consists
of time-division multiple access (TDMA) within one cluster and frequency
division multiple access (FDMA) among different clusters, making use of the
ample bandwidth provided by the mmWave communications.

\item We formulate a joint optimization problem of maximizing the sum rate
over cellular transmit beamformers, cluster receive beamformers and D2D
resource allocations for downlink transmissions of our proposed scheme. We
obtain a closed-form solution for the D2D resource allocation problem,
thereby quantitatively revealing the impacts of cellular signal strength and
D2D link quality at each standby user for practical system design.

\item Extensive system-level simulations are performed to demonstrate the
effectiveness of the proposed scheme. Compared with non-cooperation from standby users and the existing multi-hop method, approximately 2.5x to 2x improvement in terms of the 10th percentile user rate are observed when each cellular user is helped by 9 standby users. The comparison with equal resource allocation and our proposed algorithm confirms the necessity and advantages of optimizing the D2D resource allocation in our proposed scheme. In addition, we observe that fully loading the BSs is always the optimal strategy when the number of standby users exceeds a certain threshold, which is a useful insight for system design.
\end{itemize}

\subsection{Paper Organization and Notations}

The paper is organized as follows. In Section II, we introduce the system
model, present our proposed hybrid D2D-cellular scheme and formulate the
utility maximization problem. For the formulated problem, we propose a
two-step optimization algorithm in Section III. Section IV provides
numerical results of a system-level simulation and discussions of insight.

Throughout this paper, column vectors (matrices) are denoted by boldface
lower (upper) case letters. $(\mathbf{\cdot })^{T}$ represents the transpose
of a vector (matrix), $(\mathbf{\cdot })^{\ast }$ represents the conjugate
transpose, $(\cdot) ^{{\dagger }}$ represents the pseudo inverse and $\mathbf{%
Tr}\left( \mathbf{\cdot }\right) $ represents the trace. $\mathbb{C}$
denotes the complex number set and $\mathcal{C}\mathcal{N}\left( 0,\sigma
^{2}\right) $ denotes a complex Gaussian distribution with zero mean and a
variance $\sigma ^{2}$.

\section{System Model and Proposed Hybrid D2D-Cellular Scheme}

Consider a downlink multi-cell interfering broadcast channel with $N$ base
stations (BSs) serving $K$ active users in each cell. Each BS $b$, $b\in
\left\{ 1,\cdots ,N\right\} $, is equipped with $L$ transmit antennas and
each user, no matter active or non-active, has a single receive antenna. For
the considered system model, it is widely recognized that inter-cell
interference is the main cause of the rate degradation, especially in a
dense cellular networks or for cell-edge users. In this paper, we propose a
novel outband D2D cellular scheme, to enhance the capability of encountering
the inter-cell interference for end-users.

\begin{figure}[tp]
\centering\includegraphics[width=\columnwidth]{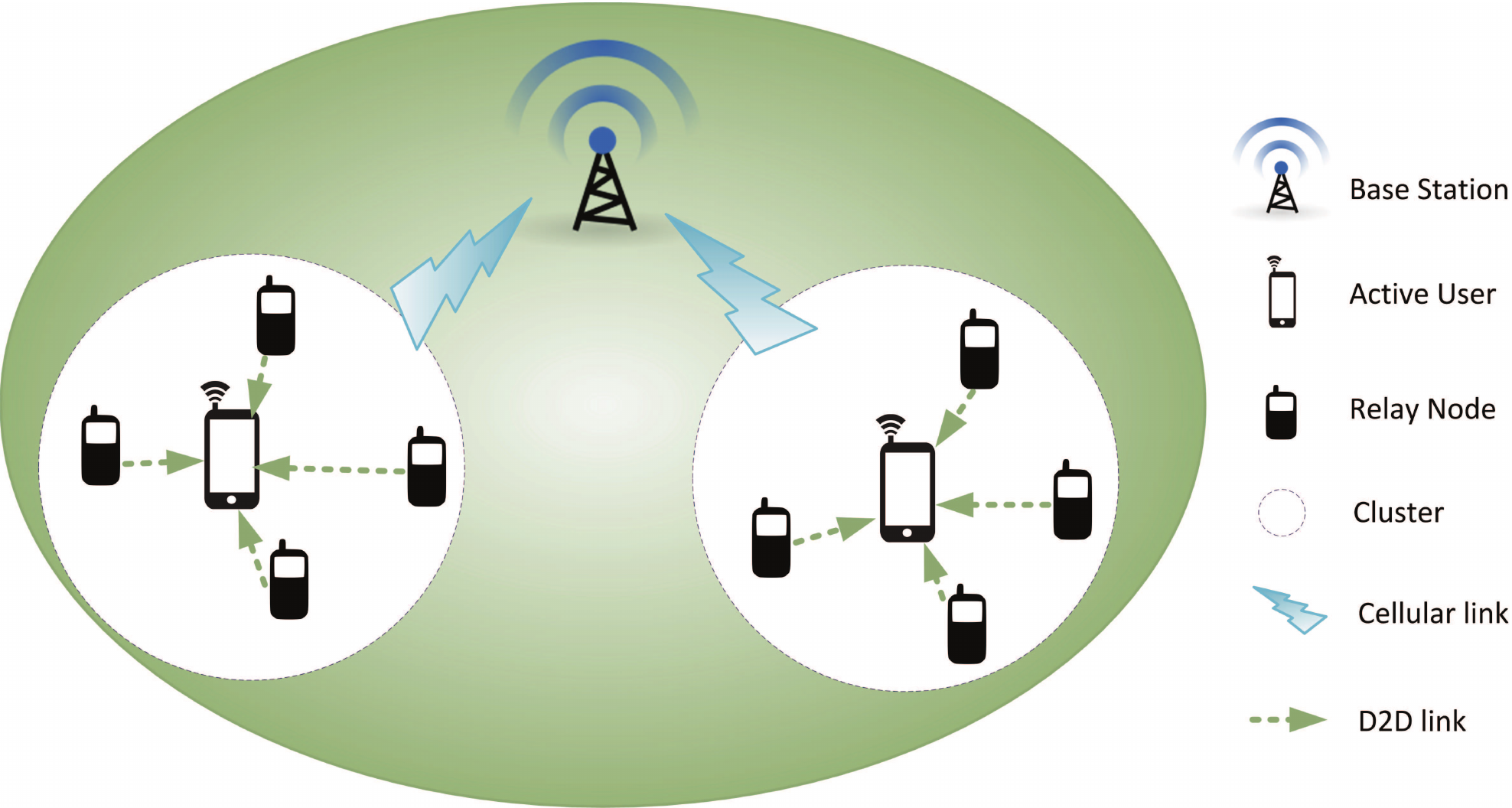} %
\centering
\caption{Proposed cellular system with outband D2D communications.}
\vspace{-5mm}
\label{fig:model}
\end{figure}

As illustrated in Fig. \ref{fig:model}, our proposed scheme consists of a
two-phase cooperative transmission procedure. In the first phase, the BSs
transmits data to active users and standby users receive signals by
\textquotedblleft overhearing\textquotedblright . In the second phase,
standby users help relay the received data to the active users so that
multiple observations of the transmitted signals are jointly processed at
the active users. The application of the distributed-MIMO technique here
enhances the capability of inter-cell interference mitigation of end users,
and thus improves the rate performance, especially for the cell-edge users.

In general, when designing a D2D communication scheme, peer discovery,
physical layer procedures and radio resource management algorithms are the
three main issues to be considered \cite{FodorM12}. We note that the peer
discovery required in our scheme is in line with conventional D2D
communication schemes. We do not focus on this problem in the paper since
existing peer discovery methods in literatures can be adopted. Instead, we
focus on physical layer procedures in this section, and then radio resource
management algorithms in Section III.

\subsection{Pre-Transmission: Clustering}

In each BS, there are three type of users: active cellular users, users
requesting D2D connections and standby users who do not request cellular nor
D2D connections. Note that the number of these standby users is usually much
larger than the number of active users. We propose a clustering strategy to
make use of these standby users to enhance the rate performance of active
users as follows.

We first define a maximum D2D communication range $d_{\max }$, which is the
maximum geographical distance between two users allowed by the user power
budget and the employed D2D communication technique. We then present some
definitions for our proposed scheme as follows.

\begin{itemize}
\item \emph{Relay node}: For each active user $k$ in each cell, all its
nearby standby users within the distance $d_{\max }$ are its potential relay
nodes. Among these nodes, those who are willing to help the active users%
\footnote{%
Users incentive for helping others has been studied in many literatures and
is not the focus of this paper.} are referred to as relay nodes. We assume
that one relay node only helps one active user at the same time. We also
assume that each BS is aware of the relay decisions of all relay nodes
within its cell.

\item \emph{Cluster}: After relaying decisions at potential relay nodes, an
active user and its relay nodes thereby form a cluster. We assume that each
active user has the same number of relay nodes for simplicity. We see that
there are $K$ clusters of identical size within each cell. We assume that
there are $M-1$ relay nodes for each active user, so that together with the
direct link from the BS to each user, an $L\times M$ MIMO channel can be
formed for each active user.
\end{itemize}

Note that this clustering strategy limits the required transmission range of
D2D links, which avoids the disadvantage of large propagation loss for long
distance transmissions when operating at the mmWave carrier frequency. We assume that cellular devices in this paper are equipped with two wireless interfaces: LTE and mmWave, so that these users have both cellular and D2D communication capabilities. This is a realistic assumption based on the development trend of RF hardware \cite{HongTAP17}.

\subsection{Phase I: BS-to-user transmission}

In our proposed scheme, for each user, there is one direct link from the BS
to the active user, and $M-1$ one-hop relay links. We denote the $k$th
active user in the cell of BS $b$, $k\in \left\{ 1,\cdots ,K\right\} $, by
user $b_{k}$ in this paper. We also assume that each active user has only a
single data stream for simplicity. Then, let $\mathbf{v}_{b_{k}}\in
\mathbb{C}
^{L\times 1}$ denote the beamformer that BS $b$ uses to transmit signals to
user $b_{k}$, and the transmitted signal from BS $b$ can be written as%
\begin{equation}
\mathbf{x}_{b}=\underset{k=1}{\overset{K}{\sum }}\mathbf{v}_{b_{k}}s_{b_{k}},
\label{Eq_x_v}
\end{equation}%
where $s_{b_{k}}$ is the message for user $b_{k}$ and $\mathbf{E}\left[\left|
s_{b_{k}}\right|^{2}\right] =1$. We assume that $s_{b_{k}}$ is chosen from a
Gaussian codebook and messages for different users are independent from each
other and from receiver noises. Collecting all beamformers $\mathbf{v}%
_{b_{k}}$ used by BS $b$, we obtain the beamformer matrix $\mathbf{V}_{b}=%
\left[ \mathbf{v}_{b_{1}},\cdots ,\mathbf{v}_{b_{K}}\right] \in
\mathbb{C}
^{L\times K}$. Then, the power budget of each BS can be presented as%
\begin{equation}
\mathbf{Tr}\left( \mathbf{V}_{b}\mathbf{V}_{b}^{\ast }\right) \leq P_{B}.
\label{Eq_PowerConst}
\end{equation}
The received signal at user $b_{k}$ can be written as%
\begin{equation}
y_{b_{k}}=\mathbf{h}_{b,b_{k}}^{T}\mathbf{v}_{b_{k}}s_{b_{k}}+\underset{%
\substack{j=1\\ j\neq k}}{\overset{K}{\sum }}\mathbf{h}_{b,b_{k}}^{T}\mathbf{%
v}_{b_{j}}s_{b_{j}}+\underset{\substack{i=1\\i\neq b}}{\overset{N}{\sum }}%
\underset{\substack{j=1\\j\neq k}}{\overset{K}{\sum }}\mathbf{h}%
_{i,b_{k}}^{T}\mathbf{v}_{i_{j}}s_{i_{j}}+n_{b_{k}},  \label{Eq_y_k}
\end{equation}%
where $\mathbf{h}_{i,b_{k}}\in \mathbb{C}^{L\times 1}$, $i\in \left\{
1,\cdots ,N\right\} $, represents the channel state information (CSI) vector
from BS $i$ to user $b_{k}$, and $n_{b_{k}}$ is the received noise at user $%
b_{k}$ and follows $\mathbb{C}\mathcal{N}\left( 0,\sigma ^{2}\right)$.

Due to the broadcasting property of the wireless channel, the relay nodes
also receive the transmitted signals from BS by \textquotedblleft
overhearing\textquotedblright . The received signals at the $m$th relay
node of user $b_{k}$, can be described as
\begin{equation}
y_{b_{k}}^{m}=\overset{N}{\underset{i=1}{\sum }}\overset{K}{\underset{j=1}{%
\sum }}\left( \mathbf{h}_{i,b_{k}}^{m}\right) ^{T}\mathbf{v}%
_{i_{j}}s_{i_{j}}+n_{b_{k}}^{m},m\in \left\{ 1,\cdots ,M-1\right\} ,
\label{Eq_Relay_Rx_}
\end{equation}%
where $\mathbf{h}_{i,b_{k}}^{m}\in
\mathbb{C}
^{L\times 1}$ is the channel coefficient from BS $i$ to the $m$th relay
node of user $b_{k}$ and $n_{b_{k}}^{m}\thicksim \mathcal{C}\mathcal{N}%
\left( 0,\sigma ^{2}\right) $ is the additive white Gaussian noise at the $m$%
th relay node of user $b_{k}$. We consider the same level of thermal noise $%
\sigma ^{2}$ at different users for simplicity.

\subsection{Phase II: Intra-cluster user cooperation}

D2D communications between users within a cluster are enabled in Phase II to realize user cooperations and ultimately facilitate BS-to-user transmissions. We assume that D2D communications take place at E-band to obtain a large available bandwidth.

\subsubsection{Operations at each relay node}

In this phase, relay nodes relay the received signals from Phase I to its
helped active users. The relaying strategy at each relay node is chosen as
follows. We note that our proposed scheme can be regarded as a general
Gaussian relay channel, in which the channel of first hop (BS-to-relay) is
weak (low signal-to-noise ratio) and the channel of second hop
(relay-to-user) is relatively strong (high signal-to-noise ratio).
Information theoretic considerations reveal that the compress-and-forward
strategy is appropriate for such a relay channel \cite{GamalBook}.
Therefore, each observation at the $m$th relay node of user $b_{k}$, $%
y_{b_{k}}^{m}$, is compressed and then forwarded to their active users.

We assume that the compression at each relay node is performed independently
to simplify the intra-cluster cooperation. The compression procedure is
modeled as the following forward test channel%
\begin{equation}
\widetilde{y}_{b_{k}}^{m}=y_{b_{k}}^{m}+e_{b_{k}}^{m},
\end{equation}%
where $\widetilde{y}_{b_{k}}^{m}$ represents the compressed signal and $e_{b_{k}}^{m}\in
\mathbb{C}
$ is the quantization noise independent of $y_{b_{k}}^{m}$. The quantization noise $e_{b_{k}}^{m}$ is also assumed
to be Gaussian distributed with zero mean and variance $q_{b_{k}}^{m}$.
After performing this compression at relay node $m$, the corresponding
information rate of $\widetilde{y}_{b_{k}}^{m}$ is%
\begin{equation}
r_{b_{k}}^{m}=\log \left( 1+\frac{\beta _{b_{k}}^{m}}{q_{b_{k}}^{m}}\right)
\text{,}  \label{Eq_C}
\end{equation}%
where
\begin{equation}
\beta _{b_{k}}^{m}=\overset{N}{\underset{i=1}{\sum }}\overset{K}{\underset{%
j=1}{\sum }}\left( \mathbf{h}_{i,b_{k}}^{m}\right) ^{T}\mathbf{v}_{i_{j}}%
\mathbf{v}_{i_{j}}^{\ast }\left( \left( \mathbf{h}_{i,b_{k}}^{m}\right)
^{T}\right) ^{\ast }+\sigma ^{2}.
\end{equation}%
We can regard $\beta _{b_{k}}^{m}$ as a parameter representing the contribution of $m$th relay node to user $b_{k}$'s achievable rate, which is defined later in (\ref{Eq_Rbk}). Smaller $\beta
_{b_{k}}^{m}$ leads to a lower information rate of the compressed signal $%
\widetilde{y}_{b_{k}}^{m}$, while smaller quantization noise $q_{b_{k}}^{m}$
leads to a higher compression rate.

Now let us denote by $t_{c}$ the frame duration of cellular transmissions
and by $B_{c}$ the cellular bandwidth. Then, during one frame duration, the
required number of information bits to be transmitted from the $m$th relay
node to user $b_{k}$ is $t_{c}B_{c}r_{b_{k}}^{m}$.

\subsubsection{Multiple access protocol}

We now consider the multiple access aspect of multiple relay nodes
transmitting to their targeted cellular user. We note that for mmWave
communications in a dense network, the angles of arrive (AOAs) of signals
from different relay nodes to one destination user might be rather close,
and these signals would interfere with each other if they are transmitted
simultaneously. Such interference can happen within one cluster as well as
between different clusters. For example, relay nodes may interfere an
unspecified cellular user when their AOAs are close to the AOAs of the
intended signals at this cellular user.

Thus, the interference between simultaneous D2D transmissions must be
managed. One possible solution here is to consider non-orthogonal multiple
access (NOMA) from relay nodes to the destination user. However, the
corresponding power allocation algorithm would be very complicated due to
the large number of relay nodes. In addition, it has been widely recognized
that successive interference cancellation (SIC) is required at the receiver
for realizing NOMA, which would increase the implementation complexity at
end-users. Furthermore, the error propagation issue of SIC is usually
difficult to deal with \cite{WeiTCom17}. For this reason, this paper adopts
orthogonal multiple access to avoid interference and to simplify the
operations at end-users.

\begin{figure}[tp]
\centering\includegraphics[width=\columnwidth]{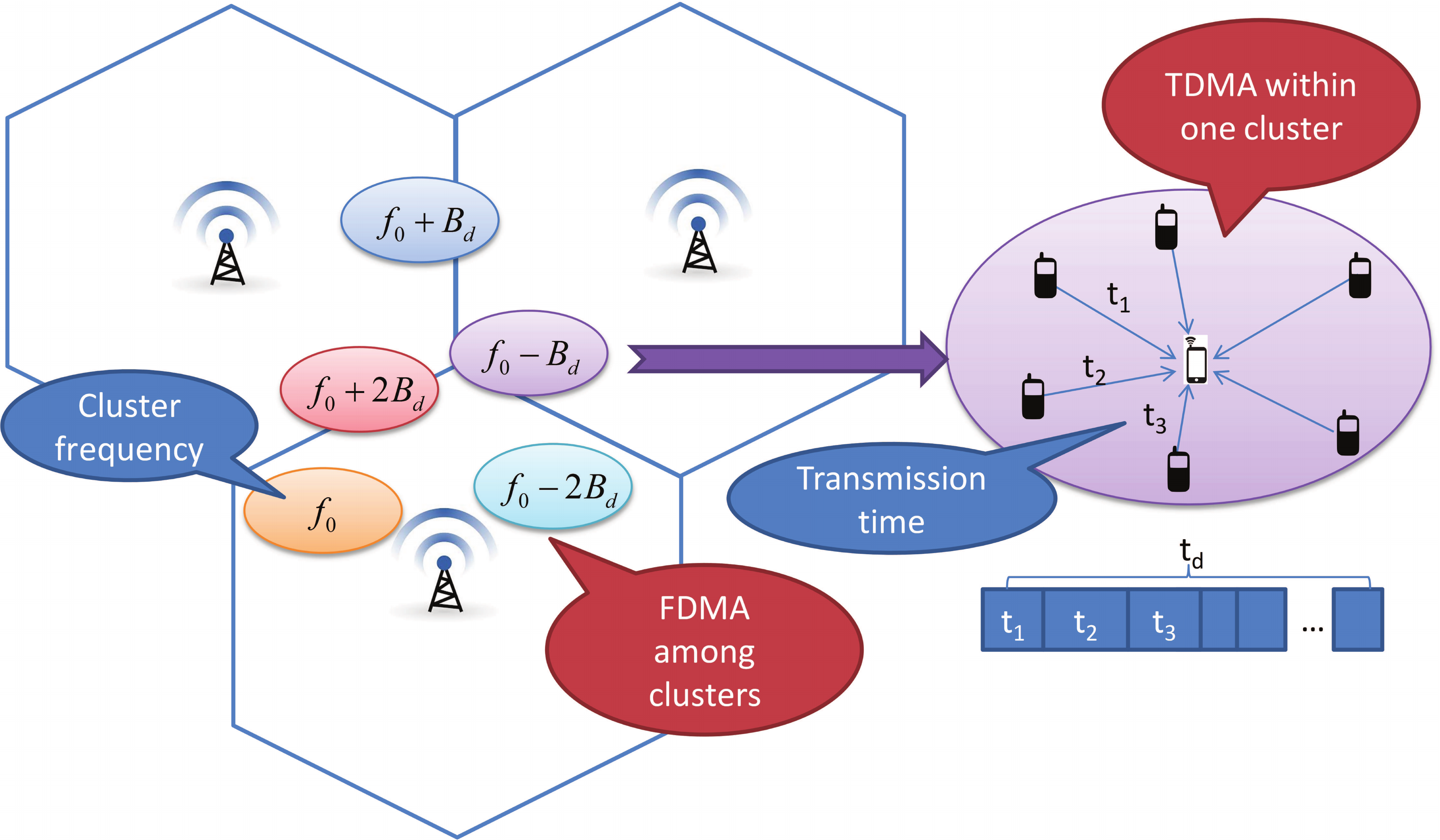} \centering
\caption{Proposed multiple access of D2D links. Here, $f_{0}$ is the carrier
frequency of D2D links, and $B_{d}$ is the bandwidth shared by one cluster.
We propose FDMA among different clusters and TDMA within one cluster.}
\label{fig:D2DAccess}
\end{figure}

\begin{figure}[tp]
\centering\includegraphics[width=0.5\columnwidth]{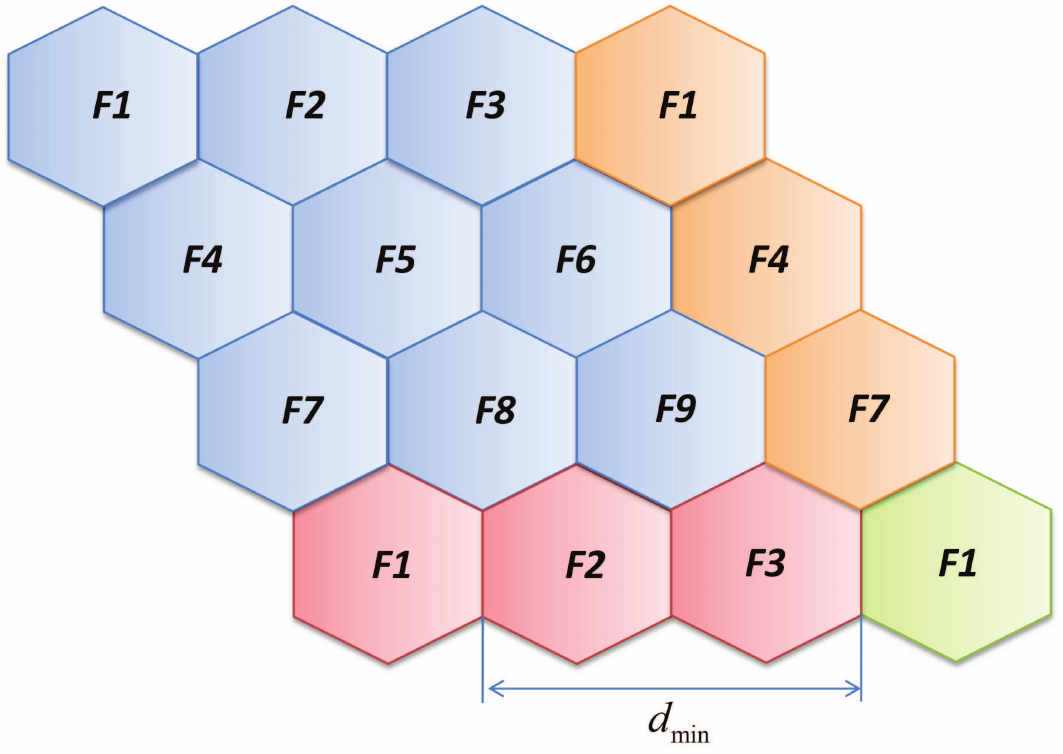} \centering
\caption{Frequency reuse pattern of the proposed FDMA protocol.}
\label{fig:FDMA_Reuse}
\end{figure}

We propose orthogonal D2D transmissions as follows. As depicted in Fig. \ref%
{fig:D2DAccess}, we adopt FDMA among different clusters and TDMA within one
cluster.

Among different clusters, we assume that the same bandwidth of $B_{d}$ is
allocated to each cluster. Thus, a bandwidth of $KB_{d}$ is required for
serving all $K$ active users within one cell. To deal with D2D interference
from adjacent cells, a multi-cell frequency reuse model is used as depicted
in Fig. \ref{fig:FDMA_Reuse}.\footnote{%
Denote the minimum distance between the sectors using the same frequency by $%
d_{\text{min}}$. Using a common path loss model in \cite{RuiTCom17,D2DModel}
and a power budget $20$ dBm, we calculate that as long as $d_{\text{min}}$
is larger than 940 m, the D2D inter-cell interference is less than 5\% of
noise power (noise power spectral density $N_{0}=169$ dBm/Hz). From Fig. \ref%
{fig:FDMA_Reuse}, we see that $d_{\text{min}}$ larger than 940 m
corresponds to a cell radius no less than 235 m. This fits the current network setups. Therefore, with this frequency reuse pattern, the
inter-cell interference is negligible in determining the rate
performance.} With this frequency reuse pattern, the total required
bandwidth is $9KB_{d}$. This design makes use of the ample bandwidth
exhibited by the mmWave communications. For example, at the E-band carrier
frequency, a total of $10$ GHz is available. We may allocate a typical $200$
MHz bandwidth for each user allowing around $50$ FDMA clusters, which is
capable of accommodating $5-6$ active cellular users within each cell.

Since FDMA is employed, all clusters are allowed to
transmit simultaneously. Let us assume that the same time duration $t_{d}$ is
available for each cluster in Phase II. Within each cluster, each relay node
is allocated with an exclusive transmission duration, which is the D2D
resource allocation parameter to be optimized. In this manner, interferences
between D2D links are thoroughly avoided.

Denote by $t_{b_{k}}^{m}$ the allocated time duration to the D2D link from
relay node $m$ to user $b_{k}$. The D2D link constraint can then be
described as%
\begin{equation}
\underset{m=1}{\overset{M-1}{\sum }}t_{b_{k}}^{m}\leq t_{d}\text{,\ \ }
\forall b,k.  \label{Eq_td}
\end{equation}%
Assuming a power budget $P_{D}$ for D2D transmissions at each relay node, we
further obtain that the actual channel capacity $C_{b_{k}}^{m}$ allowed by
the D2D channel is%
\begin{equation}
C_{b_{k}}^{m}=\log \left( 1+\frac{P_{D}\left\vert l_{b_{k}}^{m}\right\vert
^{2}}{N_{0}B_{d}}\right) \text{,}
\end{equation}%
where $l_{b_{k}}^{m}$ represents the channel coefficient of the D2D link
from relay node $m$ to user $b_{k}$, and $N_{0}$ is noise power spectral
density. Thus, we see that a total number of $%
t_{b_{k}}^{m}B_{d}C_{b_{k}}^{m} $ information bits can be transmitted via
the D2D link from relay node $m$ to user $b_{k}$.

Therefore, to guarantee no information loss through the D2D transmissions,
for the $m$th relay node of user $b_{k}$, the number of transmitted
information bits via the D2D link $t_{b_{k}}^{m}B_{d}C_{b_{k}}^{m}$ must be
larger than or equal to the required number of compression bits $t_{c}B_{c}r_{b_{k}}^{m}$%
. That is to say, the condition $t_{b_{k}}^{m}B_{d}C_{b_{k}}^{m}\geq
t_{c}B_{c}r_{b_{k}}^{m}$ must be satisfied for any $b,k,m$. In other words,
we see that the allocated time duration for the $m$th relay node of user $%
b_{k}$ must satisfy%
\begin{equation}
t_{b_{k}}^{m}\geq\frac{t_{c}B_{c}r_{b_{k}}^{m}}{B_{d}C_{b_{k}}^{m}}, \text{\
\ } \forall b,k,m.  \label{Eq_t}
\end{equation}%
The ratio $\frac{t_{c}B_{c}}{B_{d}}$ can be regarded as a constant
determined by system parameters. We see that larger the required
compression information rate $r_{b_{k}}^{m}$ and smaller the actual
channel capacity $C_{b_{k}}^{m}$, more D2D time resource is required for
relay node $m$.

\begin{rema}
Phase I and Phase II must happen consecutively without overlapping. This is
because in our design, one relay node is required to deliver information to
the active user in Phase II, after collecting all the received signals from
Phase I.
\end{rema}

\subsection{Equivalent transmission model}

\begin{figure}[tp]
\centering\includegraphics[width=0.9\columnwidth]{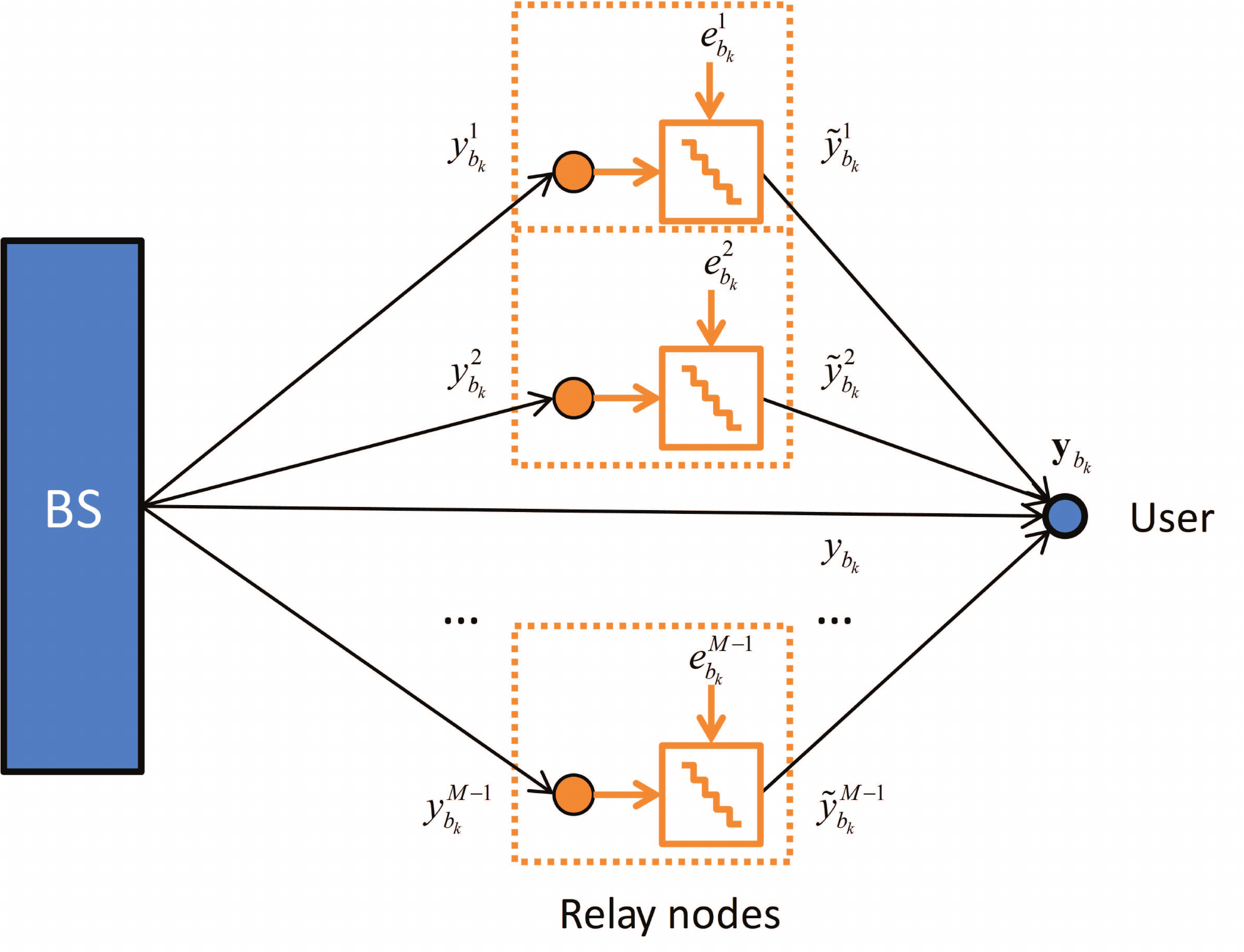} \centering
\caption{Equivalent model of the proposed scheme.}
\vspace{-5mm}
\label{fig:Equivalent_model}
\end{figure}

As illustrated in Fig. \ref{fig:Equivalent_model}, for active user $b_{k}$,
upon collecting all the compressed observations $\widetilde{y}_{b_{k}}^{m}$
from its relay nodes in Phase II and its direct observation $y_{b_{k}}$, the
received signals at user $b_{k}$ can be arranged as%
\begin{equation}
\mathbf{y}_{b_{k}}=\left[
\begin{array}{ccccc}
\widetilde{y}_{b_{k}}^{1} & \widetilde{y}_{b_{k}}^{2} & \cdots & \widetilde{y%
}_{b_{k}}^{M-1} & y_{b_{k}}%
\end{array}%
\right] ^{T}.
\end{equation}%
\qquad The equivalent transmission model from BS $b$ to user $b_{k}$ can be
further written as%
\begin{equation}
\mathbf{y}_{b_{k}}=\overset{N}{\underset{i=1}{\sum }}\overset{K}{\underset{%
j=1}{\sum }}\mathbf{H}_{i,b_{k}}\mathbf{v}_{i_{j}}s_{i_{j}}+\mathbf{n}%
_{b_{k}}+\mathbf{e}_{b_{k}},
\end{equation}%
where%
\begin{eqnarray}
\mathbf{H}_{i,b_{k}} &=&\left[
\begin{array}{ccccc}
\mathbf{h}_{i,b_{k}}^{1} & \mathbf{h}_{i,b_{k}}^{2} & \cdots & \mathbf{h}%
_{i,b_{k}}^{M-1} & \mathbf{h}_{i,b_{k}}%
\end{array}%
\right] ^{T}\in
\mathbb{C}
^{M\times L}, \notag \\
\mathbf{n}_{b_{k}} &=&\left[
\begin{array}{ccccc}
n_{b_{k}}^{1} & n_{b_{k}}^{2} & \cdots & n_{b_{k}}^{M-1} & n_{b_{k}}%
\end{array}%
\right] ^{T}\in
\mathbb{C}
^{M\times 1},  \notag %
\end{eqnarray}
and%
\begin{eqnarray}
\mathbf{e}_{b_{k}}&=& \left[
\begin{array}{ccccc}
e_{b_{k}}^{1} & e_{b_{k}}^{2} & \cdots & e_{b_{k}}^{M-1} & 0%
\end{array}%
\right] ^{T}\in
\mathbb{C}
^{M\times 1}. \notag
\end{eqnarray}%
We see that $\mathbf{n}_{b_{k}}$ is a collection of $M$ i.i.d. Gaussian
random variables and $\mathbf{n}_{b_{k}}\sim \mathcal{C}\mathcal{N}\left(
0,\sigma ^{2}\mathbf{I}_{M}\right) $. The vector $\mathbf{e}_{b_{k}}$
collects independent quantization noises from relay nodes, whose last
element is zero since there is no quantization noise in the direct
BS-to-user transmission.

Thus, by jointly processing the compressed observations $%
\begin{array}{cccc}
\widetilde{y}_{b_{k}}^{1}, \cdots , \widetilde{y}_{b_{k}}^{M-1} &  &  &
\end{array}%
$ and the direct observation $y_{b_{k}}$, user $b_{k}$ and its relay nodes
realize a distributed-MIMO channel. Then, the achievable rate for user $%
b_{k} $ is written as%
\begin{equation}
\widetilde{R}_{b_{k}}=\log \left( 1+\mathbf{v}_{b_{k}}^{\ast }\mathbf{H}%
_{b,b_{k}}^{\ast }\mathbf{J}_{b_{k}}^{-1}\mathbf{H}_{b,b_{k}}\mathbf{v}%
_{b_{k}}\right) , \label{Eq_Rbk}
\end{equation}%
where
\begin{align}
\mathbf{J}_{b_{k}}= &\underset{\substack{j=1\\ j\neq k}}{\overset{K}{\sum }}%
\mathbf{H}_{b,b_{k}}\mathbf{v}_{b_{j}}\mathbf{v}_{b_{j}}^{\ast }\mathbf{H}%
_{b,b_{k}}^{\ast }+\underset{\substack{i=1\\i\neq b}}{\overset{N}{\sum }}%
\underset{\substack{j=1\\j\neq k}}{\overset{K}{\sum }}\mathbf{H}_{i,b_{k}}%
\mathbf{v}_{i_{j}}\mathbf{v}_{i_{j}}^{\ast }\mathbf{H}_{i,b_{k}}^{\ast
} \notag \\
& +\sigma ^{2}\mathbf{I}_{M}+\left[
\begin{array}{cc}
\mathbf{Q}_{b_{k}} & 0 \\
0 & 0%
\end{array}%
\right] ,  \label{Eq_J}
\end{align}%
and%
\begin{equation}
\mathbf{Q}_{b_{k}}=\left[
\begin{array}{cccc}
q_{b_{k}}^{1} & 0 & \cdots & \cdots \\
0 & q_{b_{k}}^{2} & 0 & \cdots \\
\cdots & 0 & \cdots & 0 \\
\cdots & \cdots & 0 & q_{b_{k}}^{M-1}%
\end{array}%
\right] .
\end{equation}%
Considering the fact that Phase I and Phase II happen consecutively, the
actual information rate of user $b_{k}$ is%
\begin{equation}
R_{b_{k}}=\frac{t_{c}}{t_{d}+t_{c}}\widetilde{R}_{b_{k}}.  \label{Eq_R}
\end{equation}

We see that by permitting adjacent users to cooperate with each other, users
located in different locations can form a virtual antenna array. This allows
the deployment of MIMO techniques to single-antenna users and enhances the
channel capacity by diversity gain.

We further notice that with multiple observations at each user, receive
beamforming is also required for our proposed scheme to better realize the
MIMO capacity. In this paper, we treat interference as noise and consider a
linear receive beamforming strategy. So the estimated message is given by%
\begin{equation}
\widehat{s}_{b_{k}}=\mathbf{u}_{b_{k}}^{\ast }\mathbf{y}_{b_{k}},
\end{equation}%
where $\mathbf{u}_{b_{k}}=\left[ u_{b_{k}}^{1},u_{b_{k}}^{2},\cdots
,u_{b_{k}}^{M}\right] \in
\mathbb{C}
^{M\times 1}$ is the receive beamformer at user $b_{k}$. We see that the
receive beamformer is also a parameter to be optimized.

\subsection{Problem Formulation}

Under the context of the multi-cell interfering broadcast channel, a popular utility maximization problem is to find the
optimal transmit and receive beamformers, so that the sum rate $\underset{b=1%
}{\overset{N}{\sum }}\underset{k=1}{\overset{K}{\sum }}R_{b_{k}}$ is
maximized. On top of this, we note that the resource allocation for D2D
links is also important for our proposed scheme. This is because from (\ref%
{Eq_C}) and (\ref{Eq_t}) we see that the quantization noise variance matrix $%
\mathbf{Q}_{b_{k}}$ is constrained by the allocated time duration $%
t_{b_{k}}^{m}$ of D2D links, which ultimately affects the user rate $%
R_{b_{k}}$. This section formulates a problem of multicell sum rate
maximization by jointly optimizing transmit beamformers, receive beamformers
and quantization noise variance matrices. We also note that the CSI knowledge in this paper includes the CSI from BSs to active users
as well as that from BSs to relay nodes.

Intuitively, there are two primary factors affecting the allocated time
duration $t_{b_{k}}^{m}$ for each relay node: one is its rate contribution
to cellular user rate $R_{b_{k}}$, brought by the cellular observation $%
y_{b_{k}}^{m}$ at relay node $m$, and the other is the D2D link quality from
relay node $m$ to user $b_{k}$. These two factors may vary at different
relay nodes due to different cellular and D2D channel realizations. It is
immediate to see that if a relay node has very weak cellular observations
and poor D2D connections, we should avoid allocating much transmission time
to it. And it is also obvious that a relay node with strong cellular
observations and good D2D connections should be allocated with a fair
amount of transmission time. However, these
extreme situations are not common. For any given cellular and D2D channel realizations, how do we quantitatively allocate a reasonable transmission time
for each relay node? This D2D resource allocation problem is important yet
challenging since there is not a straightforward relationship between the
individual cellular observation at each relay node, the D2D link quality and
the rate of the destination user.

In this paper, the network utility maximization problem is
formulated as follows:
\begin{align}
\mathcal{P}_{0}: \text{\ \ \ }
\begin{array}{l}
\underset{\mathbf{v}_{b_{k},}\mathbf{u}_{b_{k}},t_{b_{k}}^{m}}{\max }\text{ }%
\underset{b=1}{\overset{N}{\sum }}\underset{k=1}{\overset{K}{\sum }}R_{b_{k}}
\\
\text{s.t.\ \ \ }\left\{
\begin{array}{l}
\text{C1: \ \ }t_{b_{k}}^{m}\geq \frac{t_{c}B_{c}r_{b_{k}}^{m}}{%
B_{d}C_{b_{k}}^{m}},\forall m,b,k \\
\text{C2: \ \ }\underset{m=1}{\overset{M-1}{\sum }}t_{b_{k}}^{m}\leq
t_{d},\forall b,k \\
\text{C3: \ \ }\mathbf{Tr}\left( \mathbf{V}_{b}\mathbf{V}_{b}^{\ast }\right)
\leq P_{B},\forall b%
\end{array}%
\right. ,%
\end{array}%
\end{align}%
where $R_{b_{k}}$ is defined as in (\ref{Eq_R}), C1 is the information flow
constraint that accounts for lossless information passing through D2D links,
C2 is the transmission time constraint for D2D links and C3 is the BS
transmission power constraint.

We see that C1 must be satisfied with equality, i.e.,%
\begin{equation}
t_{b_{k}}^{m}=\frac{t_{c}B_{c}r_{b_{k}}^{m}}{B_{d}C_{b_{k}}^{m}},
\label{Eq_t_eq}
\end{equation}%
This is because the objective function monotonically decreases with $%
q_{b_{k}}^{m}$, which also decreases with the capacity of the D2D link $%
C_{b_{k}}^{m}$. To maximize $R_{b_{k}}$, $C_{b_{k}}^{m}$ should be as large
as possible.

To better understand the property of problem $\mathcal{P}_{0}$, we transform
it into optimizing over the quantization noise covariance $q_{b_{k}}^{m}$
instead of the allocated time $t_{b_{k}}^{m}$ as follows. Denote%
\begin{equation}
w_{b_{k}}^{m}=\frac{t_{c}B_{c}}{t_{d}B_{d}C_{b_{k}}^{m}},  \label{Eq_W}
\end{equation}%
which can be interpreted as a parameter measuring the D2D link quality. Smaller the $w_{b_{k}}^{m}$ is, better the D2D link is. Thus, using this
notation and together with (\ref{Eq_C}), we combine C1 and C2, and obtain the
new D2D link constraints as%
\begin{equation}
\left\{
\begin{array}{l}
\text{C4:\ \ }\overset{M-1}{\underset{m=1}{\sum }}w_{b_{k}}^{m}\log \left( 1+%
\frac{\beta _{b_{k}}^{m}}{q_{b_{k}}^{m}}\right) \leq 1. \\
\text{C5:\ \ }q_{b_{k}}^{m}\geq 0.%
\end{array}%
\right.
\end{equation}

Then, we obtain the following problem:
\begin{eqnarray}
\mathcal{P}_{1}: \text{\ \ \ }
\begin{array}{l}
\underset{\mathbf{v}_{b_{k},}\mathbf{u}_{b_{k}},q_{b_{k}}^{m}}{\max }\text{ }%
\underset{b=1}{\overset{N}{\sum }}\underset{k=1}{\overset{K}{\sum }}R_{b_{k}}
\\
\text{s.t.\ \ \ C3, C4, C5.}%
\end{array}%
\end{eqnarray}%
We see that $\mathcal{P}_{1}$ is not a convex problem. The major difficulty
in solving $\mathcal{P}_{1}$ arises from the fact that the objective
function and the constraints are both concave in transmit beamformer $%
\mathbf{v}_{b_{k}}$ and convex in quantization noise covariance $%
q_{b_{k}}^{m}$.

\section{Enhanced Rate-performance of Cellular Users}

In this section, we propose a two-step iterative optimization method for the
formulated problem $\mathcal{P}_{1}$. In the following, we show that for
given beamformers $\mathbf{v}_{b,k}$ and $\mathbf{u}_{b,k}$, the optimization
of the quantization noise covariance $q_{b_{k}}^{m}$ can be transformed into
a convex problem and solved in closed-form by applying the
Karush-Kuhn-Tucker (KKT) conditions. Then, the optimization of beamformers $%
\mathbf{v}_{b_{k}}$ and $\mathbf{u}_{b_{k}}$ under fixed quantization noise
covariance $q_{b_{k}}^{m}$ can be regarded as a general multi-cell sum rate
maximization problem. A coordinate descent algorithm is used by applying an iterative optimization between the two steps, leading to an appropriate use of D2D links.

\subsection{Optimized D2D resource allocation}

We first assume that all transmit and receive beamformers are fixed. Then,
for given beamformers $\mathbf{v}_{b_{k}}$ and $\mathbf{u}_{b_{k}}$, the
problem of quantization noise covariance optimization at the relay node for
each single-antenna user can be described as follows:
\begin{eqnarray}
\mathcal{P}_{2}: \text{\ \ \ }
\begin{array}{l}
\underset{q_{b_{k}}^{m}}{\max }\text{ }R_{b_{k}} \\
\text{s.t.\ \ \ C4, C5.}%
\end{array}%
\end{eqnarray}%
Problem $\mathcal{P}_{2}$ is a non-convex problem and we transform it
to a convex problem as follows. Note that $\mathcal{P}_{2}$ is equivalent to
minimizing the mean square error (MSE) since a single user rate is
considered. Thus, instead of solving $\mathcal{P}_{2}$, we study the
following problem:
\begin{eqnarray}
\mathcal{P}_{3}: \text{\ \ \ }
\begin{array}{l}
\underset{q_{b_{k}}^{m}}{\min }\text{ }MSE_{b_{k}} \\
\text{s.t.\ \ \ C4, C5,}%
\end{array}%
\end{eqnarray}%
where
\begin{eqnarray}
\text{ }MSE_{b_{k}} &=&E\left[ \left( \widehat{s}_{b_{k}}-s_{b_{k}}\right)
\left( \widehat{s}_{b_{k}}-s_{b_{k}}\right) ^{\ast }\right] \notag \\
&=&\mathbf{u}_{b_{k}}^{\ast }\left( \underset{i=1}{\overset{N}{\sum }}%
\underset{j=1}{\overset{K}{\sum }}\mathbf{H}_{i,b_{k}}\mathbf{v}_{i_{j}}%
\mathbf{v}_{i_{j}}^{\ast }\mathbf{H}_{i,b_{k}}^{\ast }+\sigma ^{2}\mathbf{I}%
_{M}\right) \mathbf{u}_{b_{k}} \notag \\
&&+\mathbf{u}_{b_{k}}^{\ast }\left[
\begin{array}{cc}
\mathbf{Q}_{b_{k}} & 0 \\
0 & 0%
\end{array}%
\right] \mathbf{u}_{b_{k}}-2\func{Re}\left\{ \mathbf{u}_{b_{k}}^{\ast }%
\mathbf{H}_{b,b_{k}}\mathbf{v}_{b_{k}}\right\}  \notag \\
&=&\overset{M-1}{\underset{m=1}{\sum }}q_{b_{k}}^{m}u_{b_{k}}^{m\ast
}u_{b_{k}}^{m}+Const.  \label{Eq_MSE}
\end{eqnarray}
Then, by substituting (\ref{Eq_C}) into (\ref{Eq_MSE}), we have
\begin{eqnarray}
\mathcal{P}_{4}: \text{\ \ \ }
\begin{array}{l}
\underset{r_{b_{k}}^{m}}{\min }\text{ }\overset{M-1}{\underset{m=1}{\sum }}%
\frac{\beta _{b_{k}}^{m}u_{b_{k}}^{m\ast }u_{b_{k}}^{m}}{2^{r_{b_{k}}^{m}}-1}
\\
\text{s.t. }\overset{M-1}{\underset{m=1}{\sum }}w_{b_{k}}^{m}r_{b_{k}}^{m}%
\leq 1\text{ \& }r_{b_{k}}^{m}\geq 0\text{.}%
\end{array}%
\end{eqnarray}%
We see that the above problem is a convex optimization problem, since the
constraint is linear and the objective function is convex, which can be
verified by taking the second derivative in $r_{b_{k}}^{m}$. Now introducing
Lagrange multiplier $\mu $ and $\mathbf{\lambda }\in
\mathbb{R}
^{M-1}$, $\mu $, $\mathbf{\lambda }\geq 0$, we form the Lagrangian%
\begin{align}
L\left( C_{b_{k}}^{m},\mu ,\mathbf{\lambda }\right) = & \overset{M-1}{%
\underset{m=1}{\sum }}\frac{\beta _{b_{k}}^{m}u_{b_{k}}^{m\ast }u_{b_{k}}^{m}%
}{2^{C_{b_{k}}^{m}}-1}+\mu \left( \overset{M-1}{\underset{m=1}{\sum }}%
w_{b_{k}}^{m}C_{b_{k}}^{m}-1\right) \notag \\ &-\overset{M-1}{\underset{m=1}{\sum }}%
\lambda _{k}C_{b_{k}}^{m}.
\end{align}%
Taking the derivative of the above with respect to $r_{b_{k}}^{m}$, we apply
the KKT condition as follows:%
\begin{equation}
\frac{\partial L}{r_{b_{k}}^{m}}\mathbf{=-}\frac{\beta
_{b_{k}}^{m}u_{b_{k}}^{m\ast }u_{b_{k}}^{m}}{\left(
2^{r_{b_{k}}^{m}}-1\right) ^{2}}2^{r_{b_{k}}^{m}}\ln 2\mathbf{+}\mu
w_{b_{k}}^{m}-\lambda _{k}=0.  \label{Eq_KKT}
\end{equation}%
Note that $\lambda _{k}=0$ whenever $r_{b_{k}}^{m}>0$. Now, the optimal $%
r_{b_{k}}^{m}$ must satisfy the D2D constraint C4 with equality, i.e.,%
\begin{equation}
\overset{M-1}{\underset{m=1}{\sum }}w_{b_{k}}^{m}r_{b_{k}}^{m}=1.
\label{Eq_CC}
\end{equation}%
This is because the objective function in $\mathcal{P}_{4}$ monotonically
decreases with $r_{b_{k}}^{m}$. Solving the condition (\ref{Eq_KKT}), we
obtain the following optimal $r_{b_{k}}^{m}$ as%
\begin{equation}
\left( r_{b_{k}}^{m}\right) ^{\ast }=\log \left( \frac{a+2+\sqrt{a^{2}+4a}}{2%
}\right) ,  \label{Eq_X}
\end{equation}%
where%
\begin{equation}
a=\frac{u_{b_{k}}^{m\ast }u_{b_{k}}^{m}\ln 2}{\mu }\cdot \frac{\beta
_{b_{k}}^{m}}{w_{b_{k}}^{m}}  \label{Eq_Q}
\end{equation}%
and $\mu $ is chosen such that (\ref{Eq_CC}) is satisfied. Recall that $\beta _{b_{k}}^{m}$ represents the cellular rate contribution
while $w_{b_{k}}^{m}$ indicates the D2D link quality. Then, the parameter $a$ here represents a quantitative and comprehensive
description of the cellular observation strength and the D2D link quality of
relay node $m$.

The insights of the above solution are threefold.
\begin{enumerate}
  \item The ratio $\frac{\beta _{b_{k}}^{m}}{%
w_{b_{k}}^{m}}$ is the key parameter when allocating D2D resources. \emph{This water-filling
like solution means that we should always allocate more time to relay nodes
with strong cellular observations and good D2D link qualities.}
  \item Based on definitions in (\ref{Eq_t_eq}) and (\ref{Eq_W}), together with (\ref{Eq_X}), we obtain the allocated time to the $m$th relay of user $b_k$ as%
\begin{equation}
t_{b_{k}}^{m}=t_{d}w_{b_{k}}^{m}r_{b_{k}}^{m} = t_{d}w_{b_{k}}^{m}\log \left( \frac{a+2+\sqrt{a^{2}+4a}}{2%
}\right) .
\end{equation}%
We notice that the cellular rate contribution parameter $\beta _{b_{k}}^{m}$
affects the allocated time duration $t_{b_{k}}^{m}$ through a log function.
\emph{The saturation of log function at a large-value variable indicates that if
the received signals at all relay nodes are strong ($\beta _{b_{k}}^{m}$
large for all $m$), the D2D link quality $w_{b_{k}}^{m}$ is the main factor
to be considered}. A naive linear allocation based on D2D link quality $w_{b_{k}}^{m}$ would be sufficiently effective under this circumstance.
  \item When the D2D link qualities $w_{b_{k}}^{m}$ are
roughly at the same level within one cluster, a relatively small $t_{b_{k}}^{m}$ means that the rate contribution $\beta _{b_{k}}^{m}$ from relay node $m$ to the
destination $b_{k}$ is rather limited. In this case, we can remove relay
node $m$ from the cluster. Thus, this solution can be used as \emph{a criterion
of selecting relay nodes} when determining user clusters.
\end{enumerate}

\subsection{Joint optimization of Tx. beamformer and Rx. beamformer}

With the optimized D2D allocation and corresponding quantization noise
variances at relay nodes, we then consider the beamformer optimization for
our proposed scheme as follows:
\begin{eqnarray}
\mathcal{P}_{5}: \text{\ \ \ }
\begin{array}{l}
\underset{\mathbf{v}_{b_{k},}\mathbf{u}_{b_{k}}}{\max }\text{ }\underset{b=1}%
{\overset{N}{\sum }}\underset{k=1}{\overset{K}{\sum }}R_{b_{k}} \\
\text{s.t.\ \ \ C3.}%
\end{array}%
\end{eqnarray}

We note that such a sum rate maximization problem has been intensely
addressed in many literatures. Especially, in the context of coordinated
beamforming at different BSs, a weighted minimum mean square error (WMMSE)
approach has been proposed to solve a similar problem \cite{WMMSE} and shown
great effectiveness though numerical experiments. However, this method has a
high computation complexity when considering a large multi-cell model with
multiple antennas at each user. Since our focus here is to study the
benefits of user cooperations in the cellular network, we do not consider
high-complexity algorithms for BS cooperations in this paper.

Instead, we assume a simple situation that each BS performs random
scheduling and zero-forcing beamforming for the active users within its own
cell. Under this assumption, the benefits of user cooperations through D2D
links alone are examined and advanced BS cooperations can be further carried
out on top of our proposed scheme.

Given the considered non-cooperative BSs, we adopt a random scheduling
strategy at each BS as follows. In each cell, $S$ out of the $K$ active
users are randomly scheduled and all $K$ users must have been scheduled once
after $\left\lceil \frac{K}{S}\right\rceil $ frame durations\footnote{%
Note that $\frac{K}{S}$ is usually set to be an integer number for
simplicity.}. Denote by $\mathcal{S}_{b,t}$ the set of scheduled users in
the cell of BS $b$ at time $t$, $t\in \left\{ 1,,\cdots ,\left\lceil \frac{K%
}{S}\right\rceil \right\} $. Then, problem $\mathcal{P}_{5}$ becomes
\begin{equation*}
\mathcal{P}_{6}:\text{\ \ \ }%
\begin{array}{l}
\forall t,\text{\ } \underset{\mathbf{v}_{b_{k},}\mathbf{u}_{b_{k}}}{\max }%
\text{ }\underset{b=1}{\overset{N}{\sum }}\underset{k=1}{\overset{K}{\sum }}%
R_{b_{k}} \\
\text{s.t.\ \ \ }\left\{
\begin{array}{l}
\text{C3,} \\
\forall b_{k}\notin \mathcal{S}_{b,t},\mathbf{v}_{b_{k}}=\mathbf{0}.%
\end{array}%
\right.%
\end{array}%
\end{equation*}%
Then, the actual rate of user $b_{k}$ is%
\begin{equation}
\overline{R}_{b_{k}}=\frac{S}{K}R_{b_{k}},
\end{equation}%
since each user is served every $\frac{K}{S}$\ frame durations.

\begin{algorithm}
\caption{Resource Allocation Algorithm}
\begin{algorithmic}[1]
\STATE Initialization:
\STATE Generate the random user scheduling sets: $\forall b$, randomly generate $\frac{K}{S}$ integer sets $\mathcal{S}_{b,t}$, $t\in
\left\{ 1,\cdots ,\frac{K}{S}\right\} $ such that $\underset{t}{\cup }%
\mathcal{S}_{b,t}=\left\{ 1,\cdots ,K\right\} $ \ and $\mathcal{S}%
_{b,t_{1}}\cap \mathcal{S}_{b,t_{2}}$,  for any
$t_{1}\neq t_{2}$.
\STATE Set $\mathbf{v}_{b_{k}}$ such that $\mathbf{Tr}\left(\mathbf{v}_{b_{k}}\mathbf{v}_{b_{k}}^{\ast}\right) = \frac{P_{B}}{S}$ and $\mathbf{v}_{b_{k}}^{T}\mathbf{v}_{b_{j}} = 0 $, $\forall b, k, j \neq k.$
\STATE Let $\mathbf{Q}_{b_{k}}=\mathbf{0}_{M} $, compute the MMSE receiver $\mathbf{u}_{b_{k}}$ according to (\ref{Eq_U}), $\forall b, k.$
\FOR{$t \gets 1$ \TO $t \gets \frac{K}{S}$ }
\REPEAT
\STATE{$\forall b, k \in \mathcal{S}_{b,t} $}
\STATE{1. Compute $\mathbf{Q}_{b_{k}}$ according to (\ref{Eq_X}) and (\ref{Eq_Q}).}
\STATE{2. Update $\mathbf{u}_{b_{k}}$ according to (\ref{Eq_U}).}
\STATE{3. Compute the achievable rate $R_{b_{k}}$ according to (\ref{Eq_R}).}
\STATE{4. Find the ZF transmit beamformer $\mathbf{v}_{b_{k}}$ under above $\mathbf{u}_{b_{k}}$, according to (\ref{Eq_V}).}
\UNTIL{convergence}
\ENDFOR
\STATE Compute the actual rate for each user: $\overline{R}_{b_{k}} = \frac{S}{K} R_{b_{k}}$, $\forall b, k$.
\end{algorithmic}
\end{algorithm}

For each scheduled user $b_{s}$, fixing all the transmit beamformers and
minimizing the MSE, we obtain the well-known MMSE receiver:%
\begin{equation}
\mathbf{u}_{b_{k}}=\mathbf{J}_{b_{k}}^{-1}\mathbf{H}_{b,b_{k}}\mathbf{v}%
_{b_{k}}.  \label{Eq_U}
\end{equation}%
Then, the equivalent channel from BS $b$ to user $b_{k}$ can be seen as $%
\widetilde{\mathbf{h}}_{b_{k}}=\left( \mathbf{u}_{b_{k}}^{\ast }\mathbf{H}%
_{b,b_{k}}\right) ^{T}$. Collecting all $\widetilde{\mathbf{h}}_{b_{k}}$ for
scheduled users of BS $b$ at time-slot $t$, we have $\widetilde{\mathbf{H}}%
_{b}=\left[ \cdots ,\widetilde{\mathbf{h}}_{b_{k}},\cdots \right] ^{T}$.
Zero-forcing transmit beamforming can be easily applied using the pseudo
inverse of $\widetilde{\mathbf{H}}_{b}$ \cite{ZF}, i.e.,%
\begin{equation}
\mathbf{V}_{b}=\widetilde{\mathbf{H}}_{b}^{{\dagger }}=\widetilde{\mathbf{H}}%
_{b}^{\ast }\left( \widetilde{\mathbf{H}}_{b}\widetilde{\mathbf{H}}%
_{b}^{\ast }\right) ^{-1}.  \label{Eq_V}
\end{equation}

We then use the following algorithm for solving $\mathcal{P}_{1}$:
First, find the optimal quantization noise covariance $%
q_{b_{k}}$ for given beamformers using the solution in Section III.A; Then, compute the optimal $\mathbf{u}_{b_{k}}$ for given transmit beamformers
and quantization noise covariances using (40), and update the transmit beamformer $\mathbf{v}_{b_{k}}$ based on (41). Iterate above two steps until convergence. By doing so, we integrate two optimization problems $\mathcal{P}_{4}$ and $\mathcal{P}%
_{5}$, and thus jointly optimize the cellular and D2D resource allocation.
We use Algorithm 1 to further illustrate our proposed resource allocation
algorithm. Note that a steady convergence of Algorithm 1 is observed in simulations.

\section{Simulation Results}

In this section, numerical simulations are conducted to show the
effectiveness of the proposed algorithms. To fully demonstrate the
inter-cell interference mitigation, we consider a 19-cell wrapped-around
network with the simulation parameters listed in Table \ref{table:1}. Each
cell is a regular hexagon with a single BS located at the center, within
which cellular users are randomly distributed as shown in Fig. \ref%
{fig:19_Cell_model}. A circle centered at the each cellular user with a
radius of $d_{max}$ illustrates the user clustering as shown in Fig. \ref%
{fig:One_Cell_model}.

The cellular wireless channel is centered at a frequency of $2$ GHz and has
a bandwidth $B_{c}=20$ MHz, following the 3GPP LTE-A standard. The mmWave
wireless channel is centered at a frequency of $73$ GHz, and has several
orthogonal sub-bands of $200$ MHz so that an exclusive bandwidth can be
occupied by each cluster, which is shared within one cluster via TDMA. The
channel from the relay nodes to each cellular user has line-of-sight (LoS), with the path
loss given by $69.7+24log10(d_{m})$ dB\cite{RuiTCom17}. In addition, we
consider a Nakagami fading with the Nakagami parameter $\alpha =4$ as
assumed in many mmWave D2D works. Each relay node is assumed to transmit at
a fixed power of $20$ dBm with an antenna gain of $27$ dBi \cite{D2DModel}.

\begin{figure}[tp]
\centering
\begin{subfigure}[t]{1.0\columnwidth}
\captionsetup{width=1.0\textwidth}
\captionsetup[sub]{font=it,labelfont={bf,it}}
        \includegraphics[width=\textwidth]{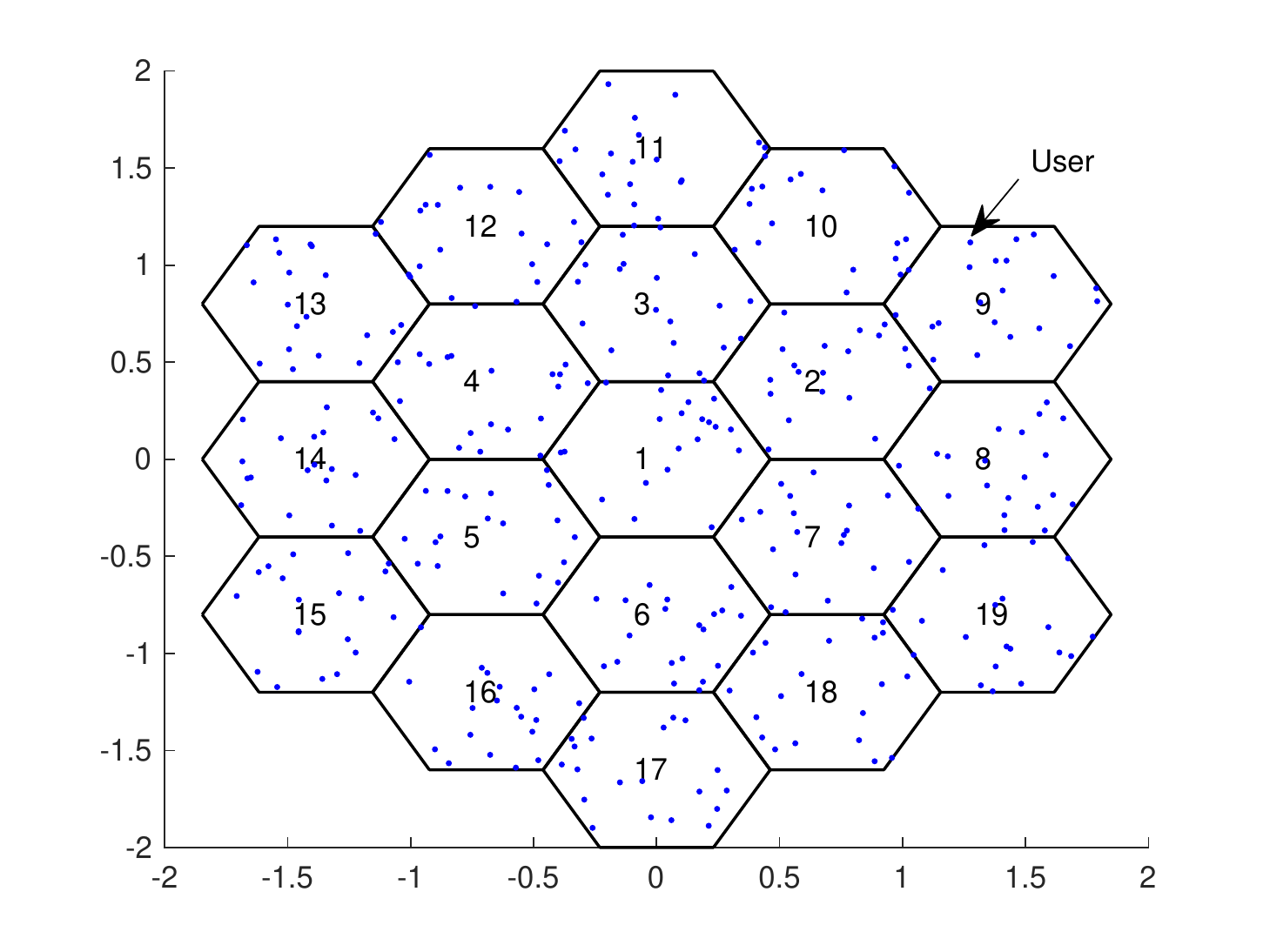}
         \caption{19-cell wrapped around model used in simulations.}
            \label{fig:19_Cell_model}
    \end{subfigure}
~
\begin{subfigure}[t]{1.0\columnwidth}
\captionsetup{width=1.0\textwidth}
\captionsetup[sub]{font=it,labelfont={bf,it}}
        \includegraphics[width=\textwidth]{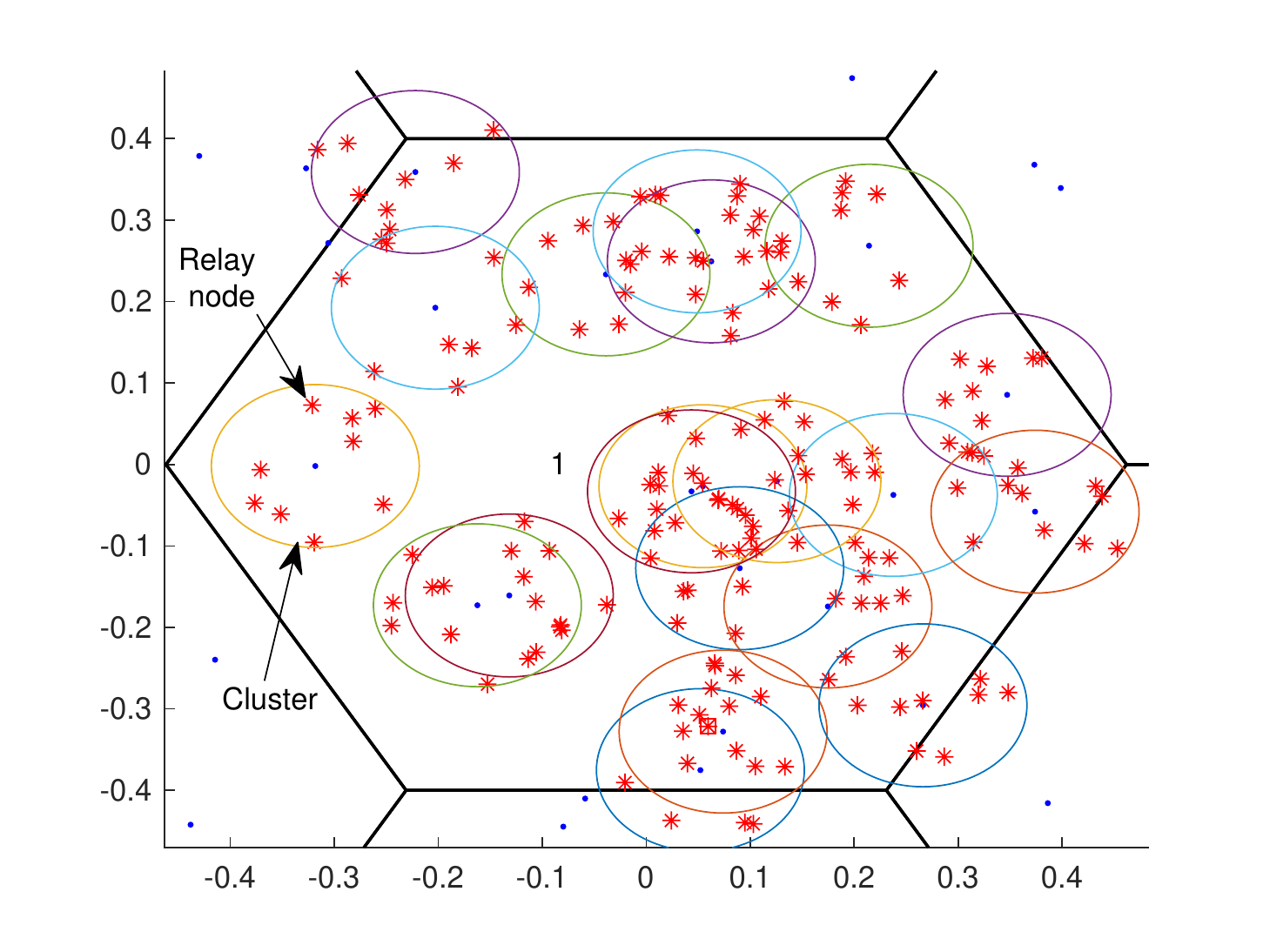}
             \caption{User clustering illustrated in one cell.}
        \label{fig:One_Cell_model}
    \end{subfigure}
\caption{19-cell wrapped around network. Dots represent the active cellular
users and stars stand for the relay nodes. Each active cellular user and its
relay nodes are circled out by an oval, representing one cluster. }
\vspace{-5mm}
\end{figure}

\newcounter{TmpEqCnt0} \setcounter{TmpEqCnt0}{\value{equation}} %
\setcounter{equation}{38}
\begin{figure*}[hb]
\normalsize \hrule%
\begin{equation}
c_{b_{k}}=\log \left( 1+\frac{\mathbf{h}_{b,b_{k}}^{T}\mathbf{v}_{b_{k}}%
\mathbf{v}_{b_{k}}^{\ast }\left( \mathbf{h}_{b,b_{k}}^{T}\right) ^{\ast }}{%
\underset{\substack{j=1\\ j\neq k}}{\overset{K}{\sum }}\mathbf{h}%
_{b,b_{k}}^{T}\mathbf{v}_{b_{j}}\mathbf{v}_{b_{j}}^{\ast }\left( \mathbf{h}%
_{b,b_{k}}^{T}\right) ^{\ast }+\underset{\substack{i=1\\i\neq b}}{\overset{N}%
{\sum }}\underset{\substack{j=1\\j\neq k}}{\overset{K}{\sum }}\mathbf{h}%
_{i,b_{k}}^{T}\mathbf{v}_{i_{j}}\mathbf{v}_{i_{j}}^{\ast }\left( \mathbf{h}%
_{i,b_{k}}^{T}\right) ^{\ast }+\sigma ^{2}}\right) , \label{Eq_R_bk_Multihop_1}
\end{equation}%
and%
\begin{flalign}
c_{b_{k}}^{m}=\log \left( 1+\frac{\left( \mathbf{h}_{b,b_{k}}^{m}\right) ^{T}%
\mathbf{v}_{b_{k}}\mathbf{v}_{b_{k}}^{\ast }\left( \left( \mathbf{h}%
_{b,b_{k}}^{m}\right) ^{T}\right) ^{\ast }}{\underset{\substack{j=1\\ j\neq k%
}}{\overset{K}{\sum }}\left( \mathbf{h}_{b,b_{k}}^{m}\right) ^{T}\mathbf{v}%
_{b_{j}}\mathbf{v}_{b_{j}}^{\ast }\left( \left( \mathbf{h}%
_{b,b_{k}}^{m}\right) ^{T}\right) ^{\ast }+\underset{\substack{i=1\\i\neq b}}%
{\overset{N}{\sum }}\underset{\substack{j=1\\j\neq k}}{\overset{K}{\sum }}%
\left( \mathbf{h}_{i,b_{k}}^{m}\right) ^{T}\mathbf{v}_{i_{j}}\mathbf{v}%
_{i_{j}}^{\ast }\left( \left( \mathbf{h}_{i,b_{k}}^{m}\right) ^{T}\right)
^{\ast }+\sigma ^{2}}\right). \label{Eq_R_bk_Multihop_2}
\end{flalign}
\end{figure*}

\setcounter{equation}{36}

We compare the performance of the following benchmark schemes with our
proposed scheme by extensive system-level simulations.

\textbf{Benchmark Scheme 1: No user cooperation.} We assume that each
single-antenna user has no relay node. This can be regarded as the
performance baseline. If the rate performance of our proposed scheme is even
worse than this benchmark, there is no need to perform user cooperation as
proposed.

\textbf{Benchmark Scheme 2: Ideal user cooperation. }Consider an ideal case
of infinite D2D link capacity, which is obtained by assuming no quantization
noise at the relay nodes, i.e., fixing $\mathbf{Q}_{b,k}=\mathbf{0}_{M}$.
This can be seen as a performance upper bound of our proposed scheme. In
this case, since an infinite D2D link capacity is assumed, having $M-1$
relay nodes for each cellular user is equivalent to that each user is
equipped with $M$ geographically dispersed receive antennas.

\textbf{Benchmark Scheme 3: Equal D2D resource allocation. }Consider an
equal D2D resource allocation, where each relay node occupies the same length
of transmission time, i.e.,
\begin{equation}
t_{b_{k}}^{m}=\frac{t_{d}}{M-1}.
\end{equation}

\textbf{Benchmark Scheme 4: Multi-hop D2D cooperation. }Consider a multi-hop
D2D cooperation scheme as in past literature \cite{WangTWC15}\cite%
{ZhangTWC09}. Within one cluster, the relay node with the strongest channel
will decode the message from BS and forward it to the user through the D2D
connection. We assume that the cellular and D2D transmissions happen
consecutively with a negligible time delay caused by establishing the relay
link. Also, the D2D link capacity is much larger than the cellular
transmission rate. Thus, the achievable rate of user $b_{k}$ is the maximum
rate within its cluster, i.e.,%
\begin{equation}
R_{b_{k}}^{\mathbf{Multi-hop}}=\max \left\{ c_{b_{k}},\underset{m}{\max }%
c_{b_{k}}^{m}\right\} \text{,}
\end{equation}%
where $c_{b_{k}}$ and $c_{b_{k}}^{m}$ are presented in (\ref{Eq_R_bk_Multihop_1}) and (\ref{Eq_R_bk_Multihop_2}) respectively.

\begin{table}[!]
\caption{SIMULATION PARAMETERS.}
\label{table:1}\centering  \renewcommand{\arraystretch}{0.8}
\begin{tabular}{|c|p{3.5cm}|}
\cline{1-2}\hline
{\small Cellular Layout} & {\small Hexagonal 19-cell wrapped-around} \\
\hline
{\small Cellular bandwidth} & $20${\small \ MHz} \\ \hline
{\small Cellular frame duration} & $1.25${\small \ ms} \\ \hline
{\small D2D bandwidth for one cluster} & $200${\small \ MHz} \\ \hline
{\small Distance between cells} & $0.8${\small \ km} \\ \hline
{\small Max. D2D transmission range} & $100$ m \\ \hline
{\small Num. of users} & ${\small 20}$ \\ \hline
{\small Max. Tx power for BSs} & $43${\small \ dBm} \\ \hline
{\small Max. Tx power for Relay nodes} & $20${\small \ dBm} \cite{D2DModel}
\\ \hline
{\small Cellular Antenna gain} & $15${\small \ dBi} \\ \hline
{\small D2D Antenna gain(mmWave)} & $27${\small \ dBi} \cite{D2DModel} \\
\hline
{\small Background noise} & $169${\small dBm/Hz} \\ \hline
{\small Path loss from BS to user} & ${\small 128.1+37.6log10(d_{km})}$ dB
\\ \hline
{\small Path loss of D2D link} & ${\small 69.7+24log10(d_{m})}$ dB\cite%
{RuiTCom17} \\ \hline
{\small Log-normal shadowing} & $8$ dB \\ \hline
{\small Rayleigh small scale fading} & $0$ dB \\ \hline
{\small Nakagami parameter of D2D links} & $4$ \\ \hline
\end{tabular}%
\end{table}

For the above benchmark schemes and our proposed scheme, Algorithm 1\footnote{%
If Benchmark Scheme 1 and 4 are considered, omit the step 1 within each
iteration.} is executed over $T$ different channel realizations to measure
the cellular rate performance by the averaged long-term rate. We have $T=100$
for the shown results in this paper.

\begin{figure}[tp]
\centering
\begin{subfigure}[t]{0.45\columnwidth}
\centering \captionsetup{width=1.0\textwidth} %
\captionsetup[sub]{font=it,labelfont={bf,it}} \includegraphics[width=%
\textwidth]{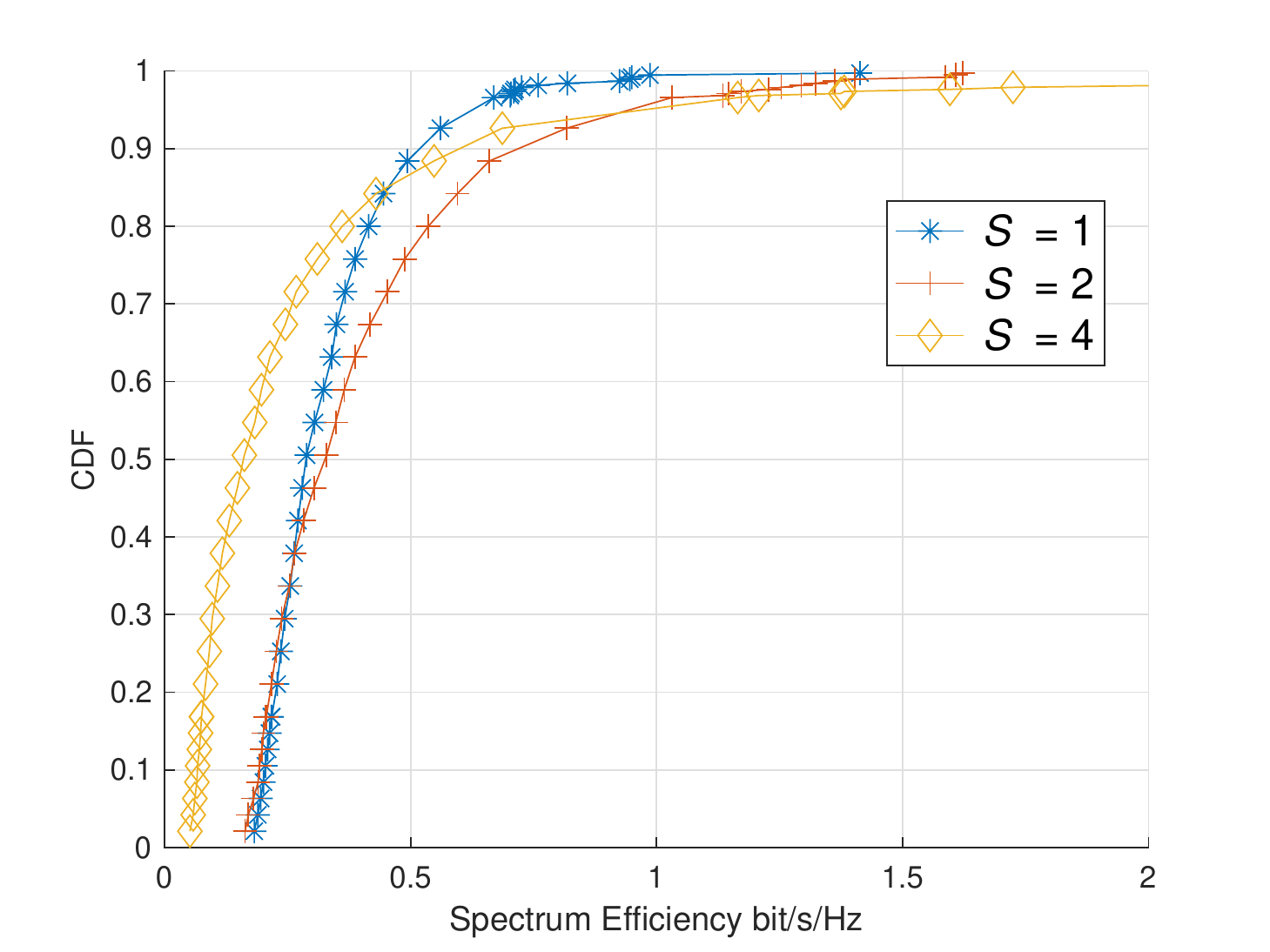}
\caption{$M$ = 1, Benchmark 1 (No user cooperation).}
\label{fig:19cells4Tx_1}
\end{subfigure}
~
\begin{subfigure}[t]{0.45\columnwidth}
\centering \captionsetup{width=1.0\textwidth} %
\captionsetup[sub]{font=it,labelfont={bf,it}} \includegraphics[width=%
\textwidth]{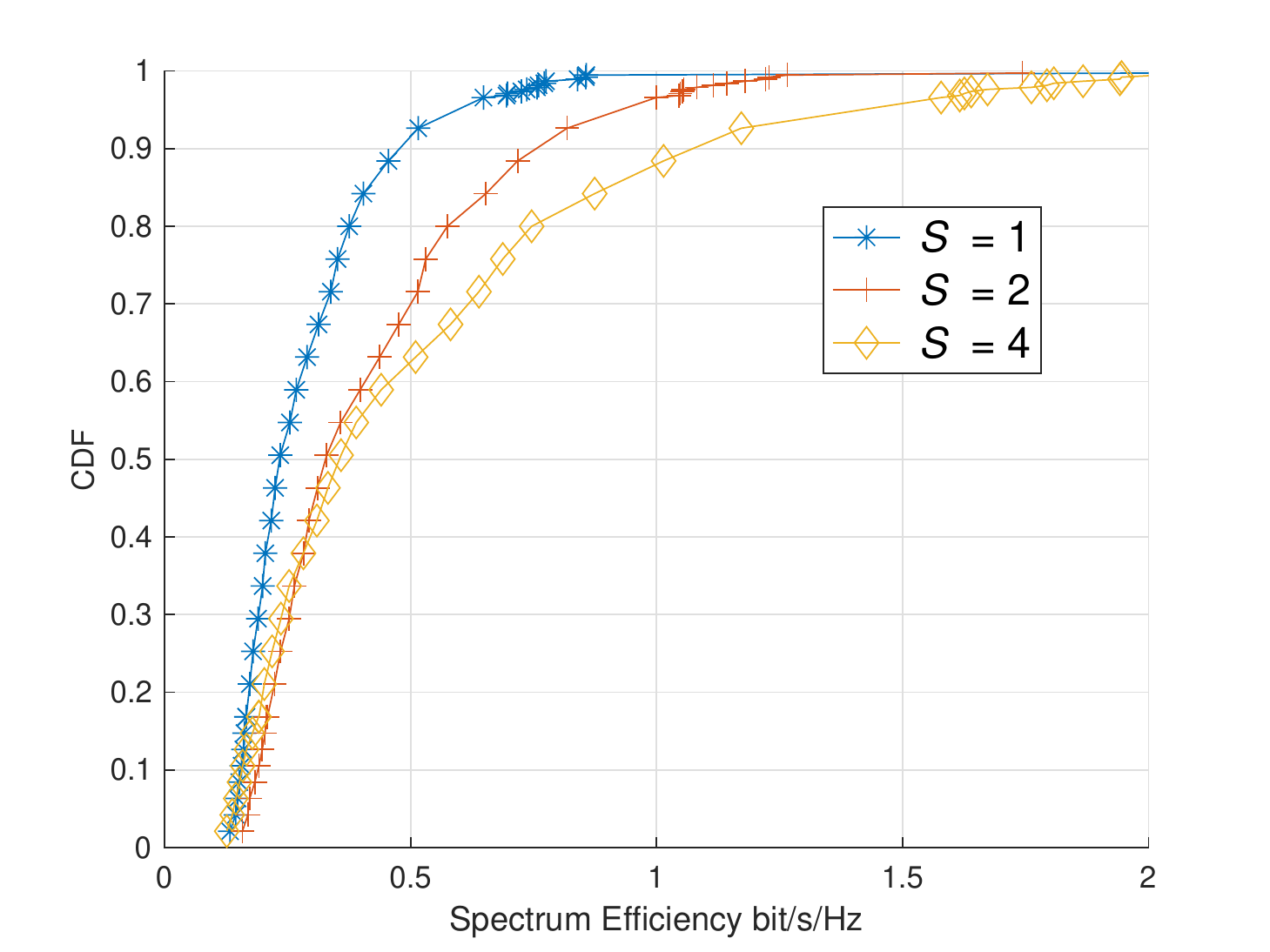}
\caption{$M$ = 2, Benchmark 2 (Ideal cooperation).}
\label{fig:19cells4Tx_2}
\end{subfigure}
~
\begin{subfigure}[t]{0.45\columnwidth}
\centering \captionsetup{width=1.0\textwidth} %
\captionsetup[sub]{font=it,labelfont={bf,it}} \includegraphics[width=%
\textwidth]{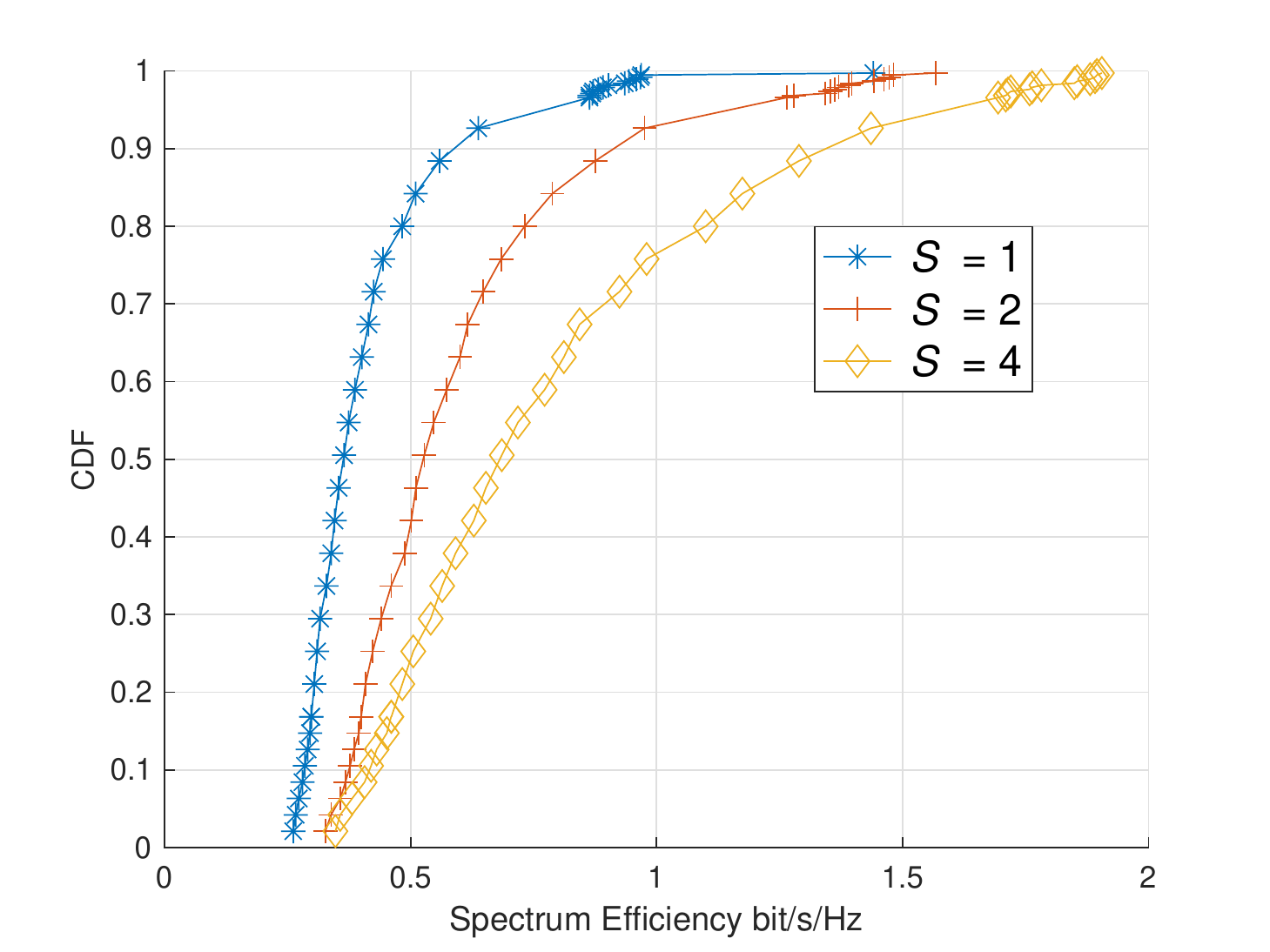}
\caption{$M$ = 5, Benchmark 2 (Ideal cooperation).}
\label{fig:19cells4Tx_5}
\end{subfigure}
~
\begin{subfigure}[t]{0.45\columnwidth}
\centering \captionsetup{width=1.0\textwidth} %
\captionsetup[sub]{font=it,labelfont={bf,it}} \includegraphics[width=%
\textwidth]{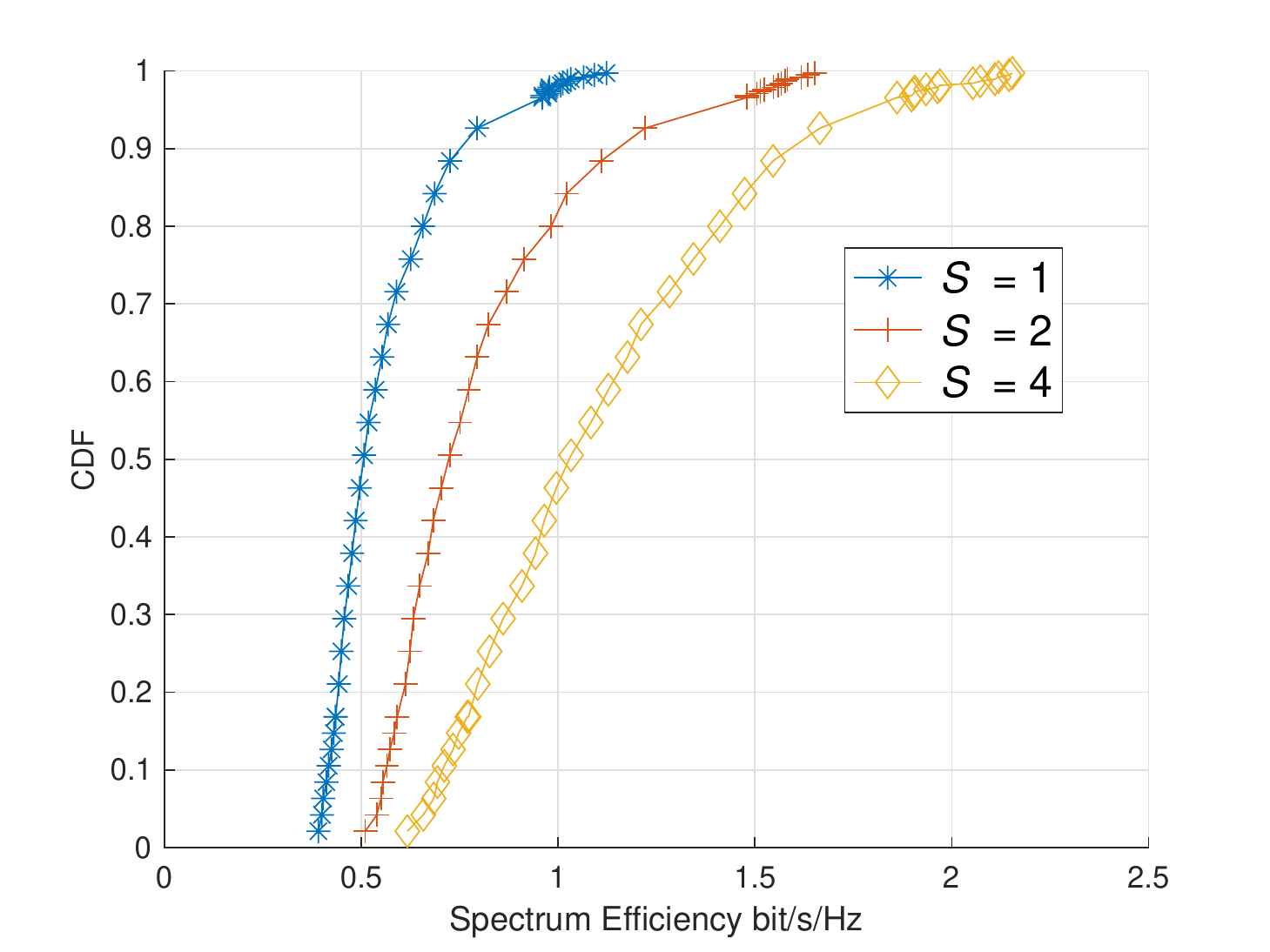}
\caption{$M$ = 10, Benchmark 2 (Ideal cooperation).}
\label{fig:19cells4Tx_10}
\end{subfigure}
\caption{Cumulative distribution function of user data rate comparison with
different number of scheduled users $S$, the number of transmit antennas at
each BS $L = 4$, 19-cell wrapped-around.}
\vspace{-5mm}
\label{fig:19cells4Tx_M}
\end{figure}

\begin{figure}[tp]
\centering
\begin{subfigure}[t]{0.45\columnwidth}
\centering \captionsetup{width=1.0\textwidth} %
\captionsetup[sub]{font=it,labelfont={bf,it}} \includegraphics[width=%
\textwidth]{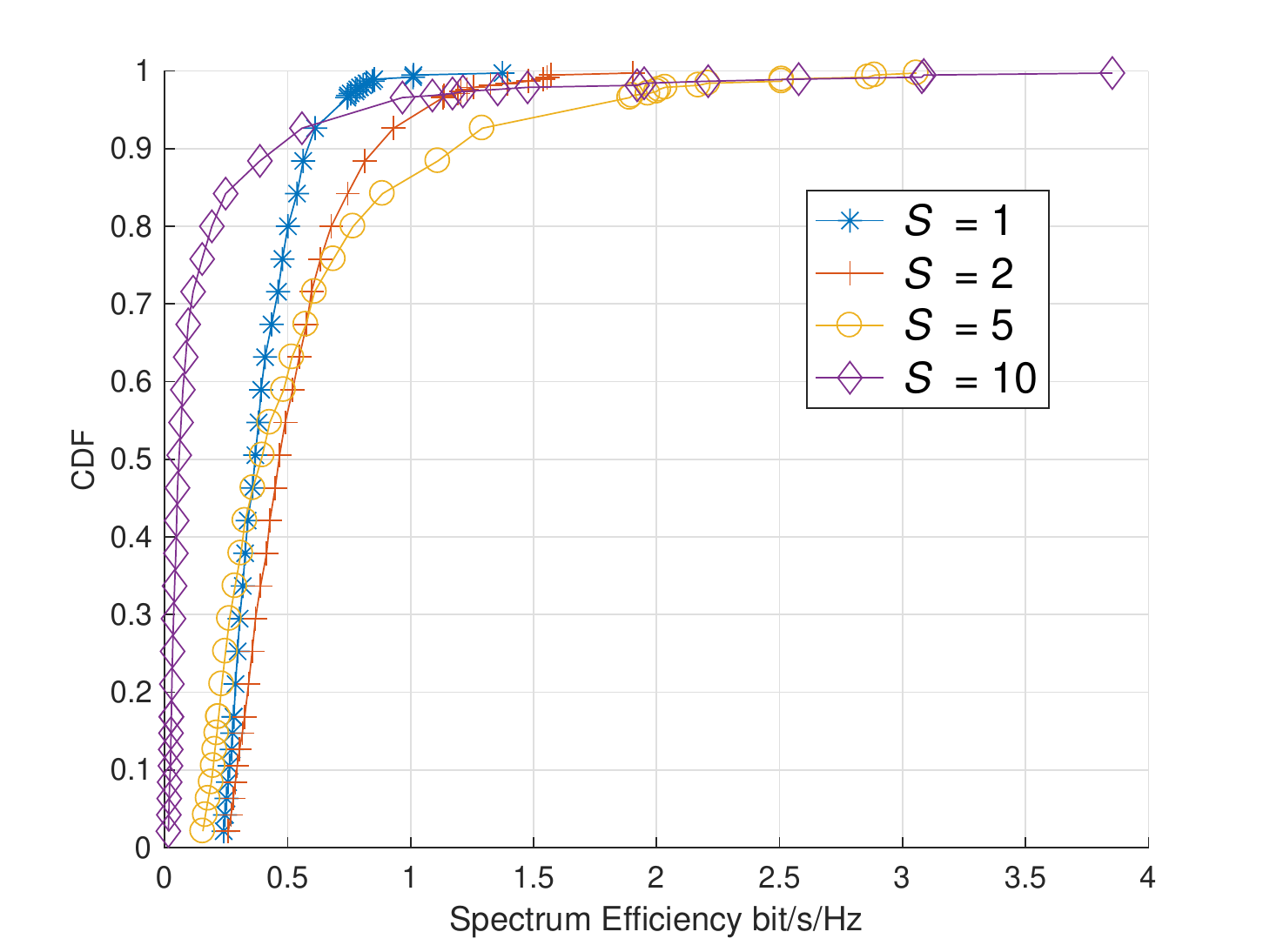}
\caption{$M$ = 1, Benchmark 1 (No user cooperation).}
\label{fig:19cells10Tx_1}
\end{subfigure}
~
\begin{subfigure}[t]{0.45\columnwidth}
\centering \captionsetup{width=1.0\textwidth} %
\captionsetup[sub]{font=it,labelfont={bf,it}} \includegraphics[width=%
\textwidth]{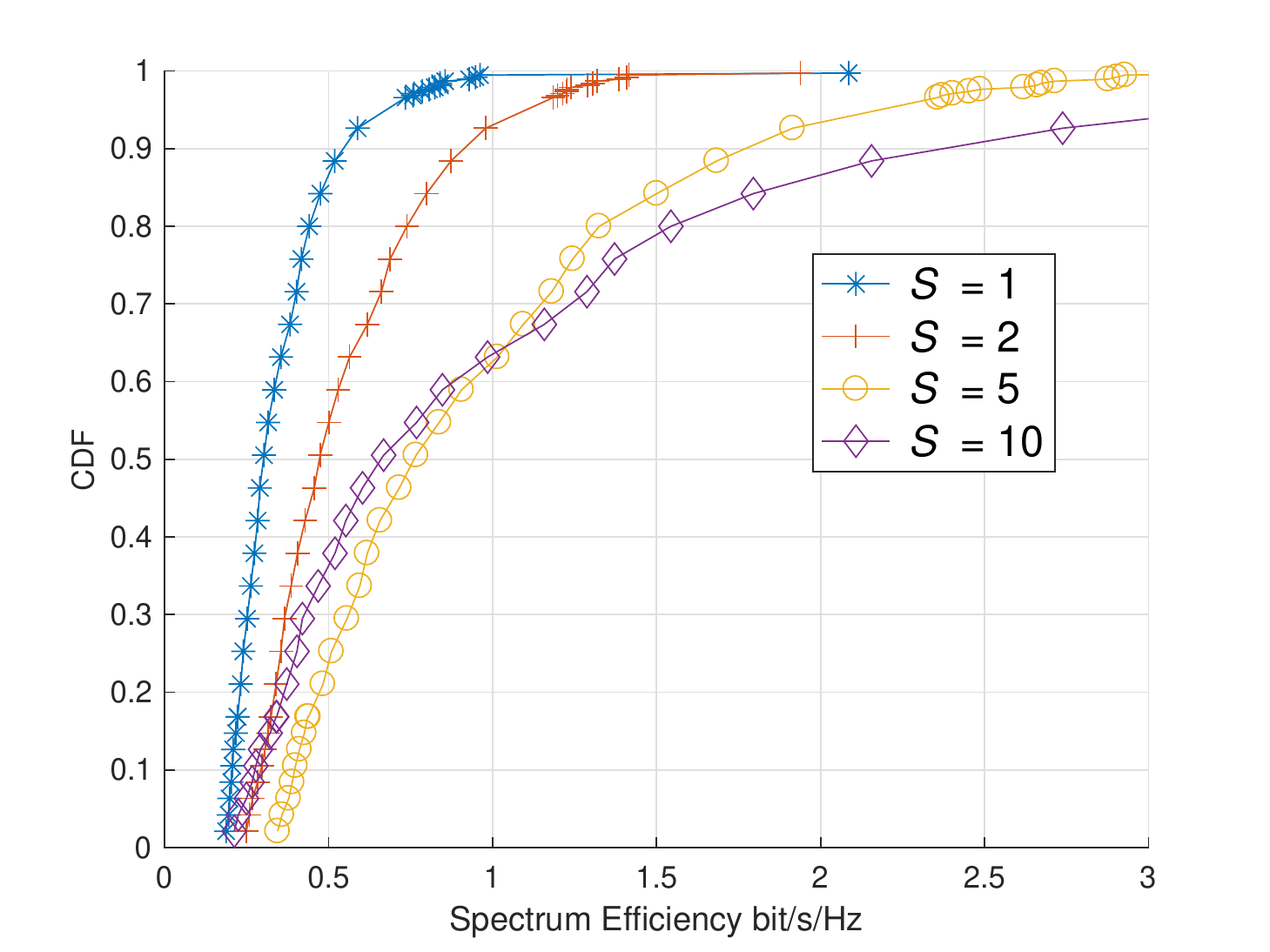}
\caption{$M$ = 2, Benchmark 2 (Ideal cooperation).}
\label{fig:19cells10Tx_2}
\end{subfigure}
~
\begin{subfigure}[t]{0.45\columnwidth}
\centering \captionsetup{width=1.0\textwidth} %
\captionsetup[sub]{font=it,labelfont={bf,it}} \includegraphics[width=%
\textwidth]{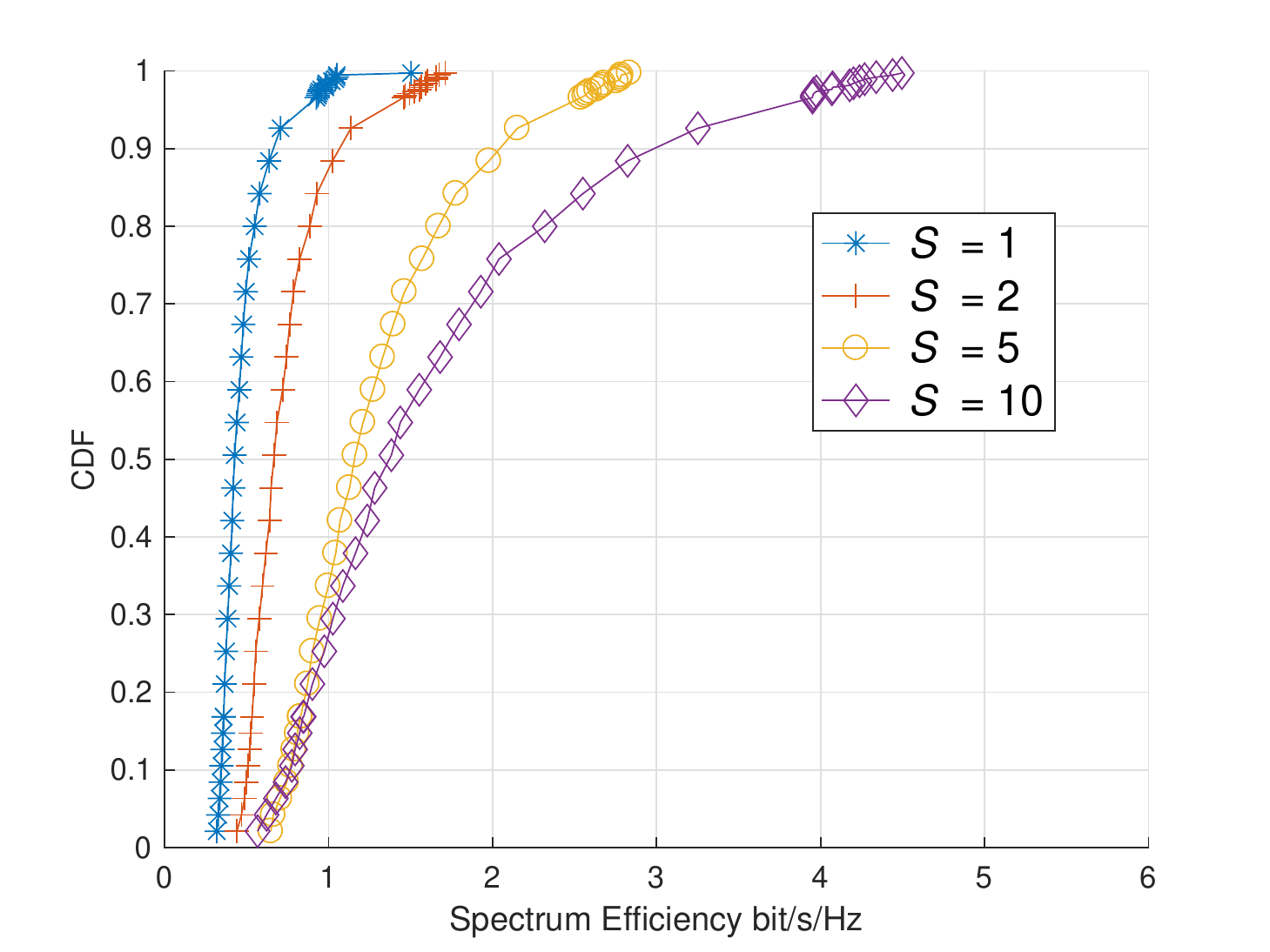}
\caption{$M$ = 5, Benchmark 2 (Ideal cooperation).}
\label{fig:19cells10Tx_5}
\end{subfigure}
~
\begin{subfigure}[t]{0.45\columnwidth}
\centering \captionsetup{width=1.0\textwidth} %
\captionsetup[sub]{font=it,labelfont={bf,it}} \includegraphics[width=%
\textwidth]{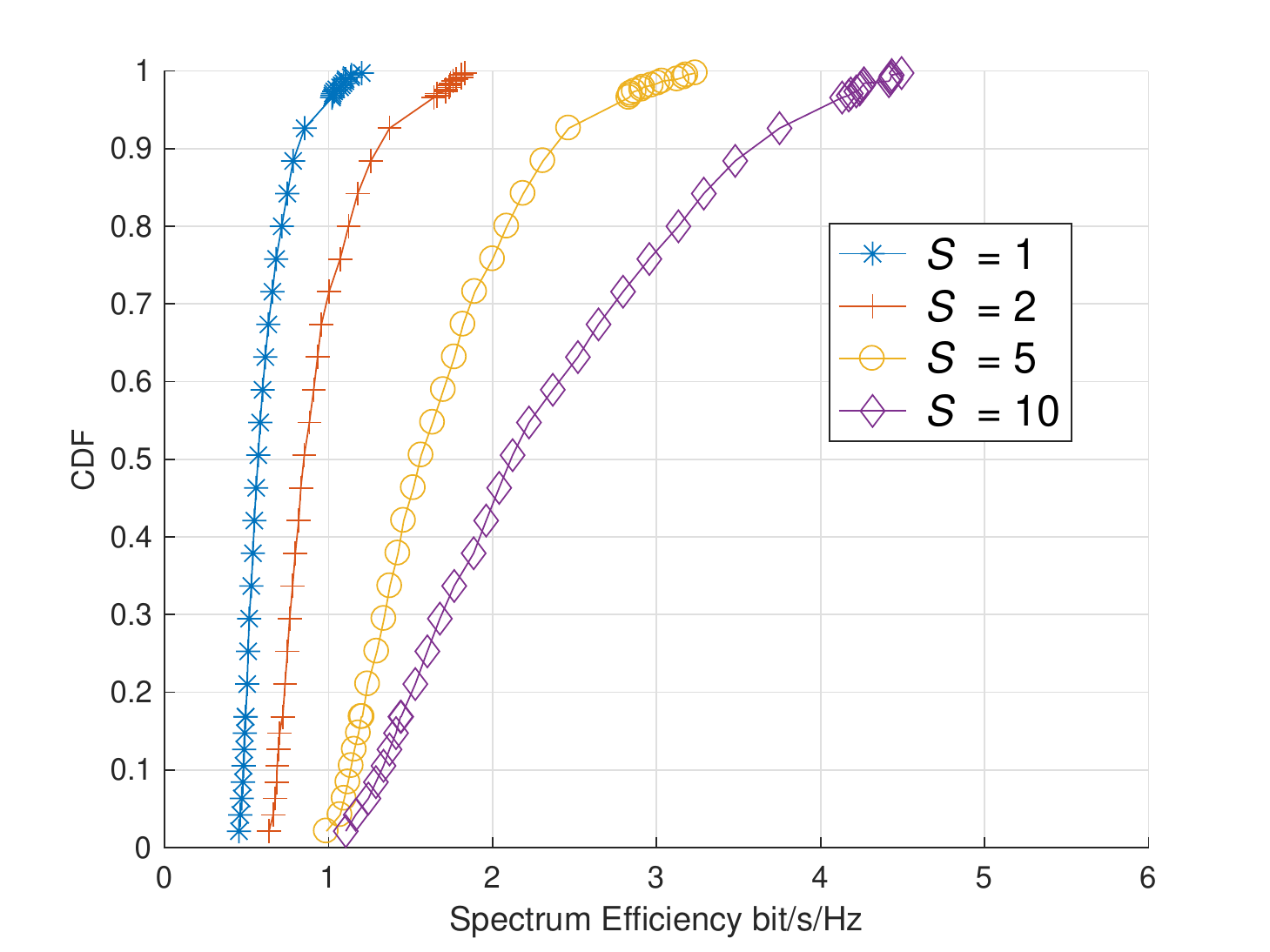}
\caption{$M$ = 10, Benchmark 2 (Ideal cooperation).}
\label{fig:19cells10Tx_10}
\end{subfigure}
\caption{Cumulative distribution function of user data rate comparison with
different number of scheduled users $S$, the number of transmit antennas at
each BS $L = 10$, 19-cell wrapped-around.}
\vspace{-5mm}
\label{fig:19cells10Tx_M}
\end{figure}

\begin{figure}[tbp]
\centering
\begin{subfigure}[t]{0.45\columnwidth}
\centering \captionsetup{width=1.0\textwidth} %
\captionsetup[sub]{font=it,labelfont={bf,it}} \includegraphics[width=%
\textwidth]{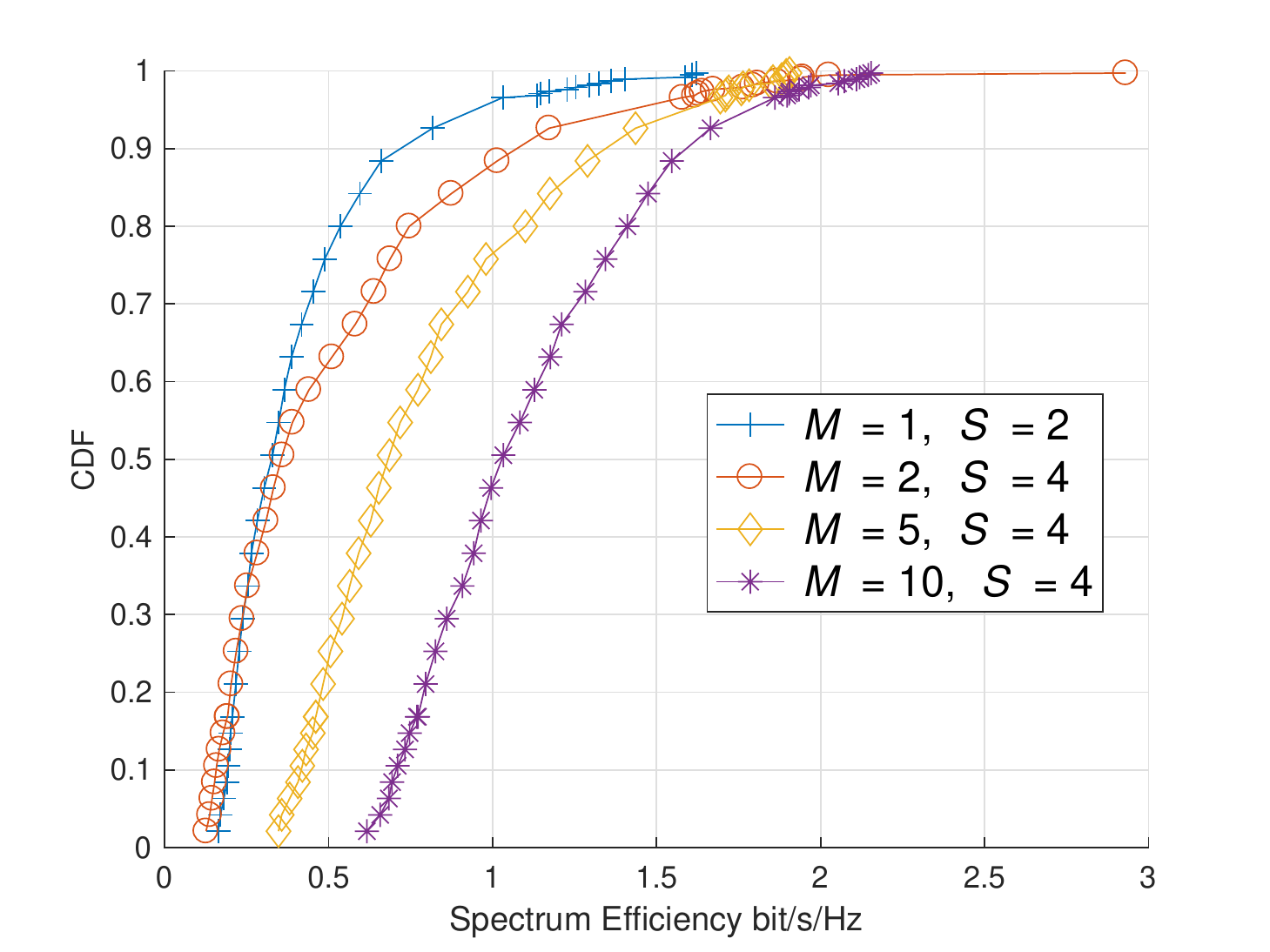}
\caption{$L = 4$, Benchmark 2 (Ideal cooperation).}
\label{fig:19cells4Tx}
\end{subfigure}
~
\begin{subfigure}[t]{0.45\columnwidth}
\centering \captionsetup{width=1.0\textwidth} %
\captionsetup[sub]{font=it,labelfont={bf,it}} \includegraphics[width=%
\textwidth]{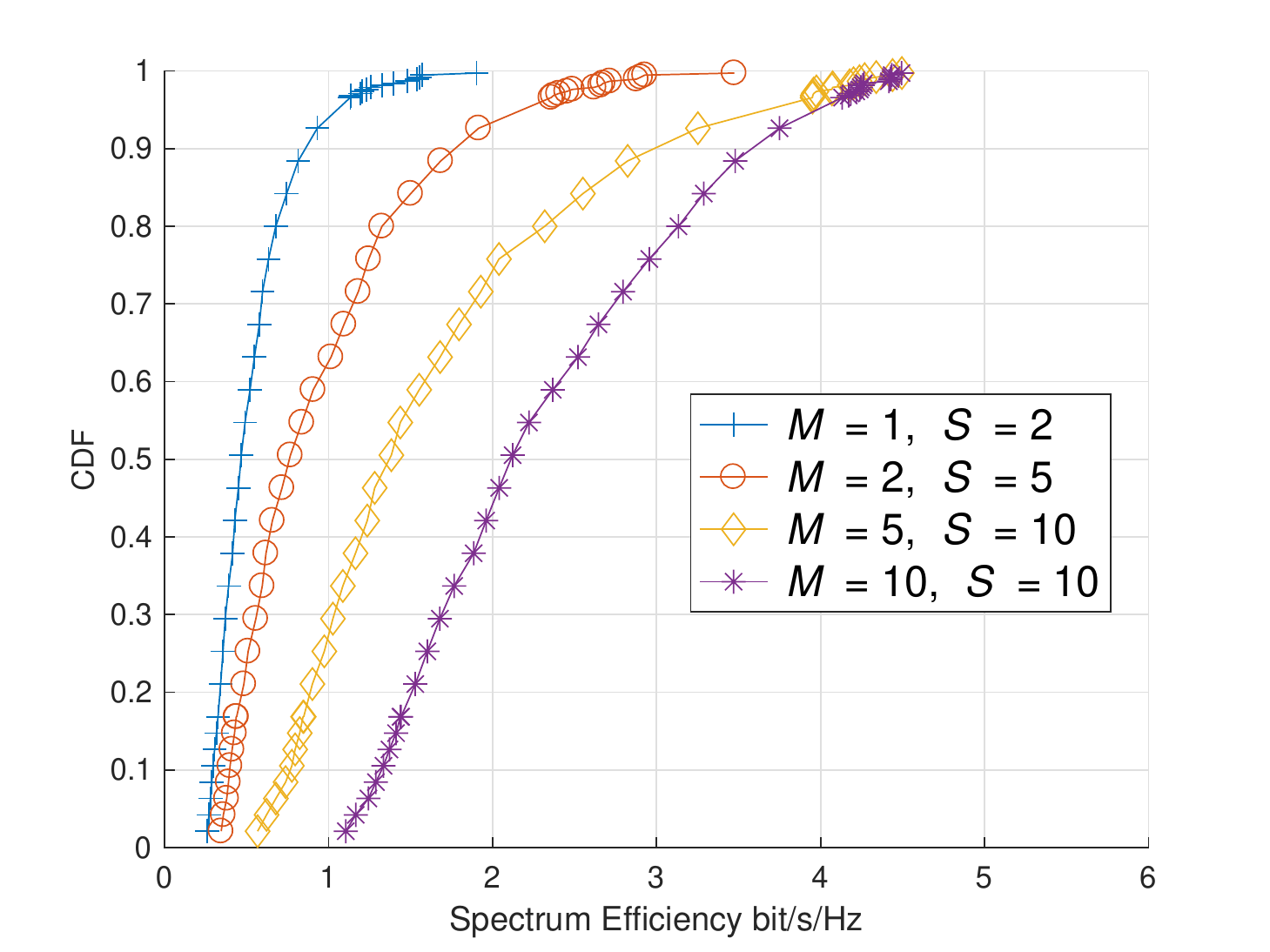}
\caption{$L = 10$, Benchmark 2 (Ideal cooperation).}
\label{fig:19cells10Tx}
\end{subfigure}
\caption{Cumulative distribution function of user data rate comparison with
different number of received antennas $M$.}
\vspace{-5mm}
\label{fig:19cellsMTx}
\end{figure}

\begin{table*}[b]
\caption{User data rate comparisons, $L = 4$.}
\label{table:2}\centering  \renewcommand{\arraystretch}{0.8}
\begin{tabular}{ccccc}
\hline
Scheme & 10th Percentile Rate (Mbps) & Gain & 50th Percentile Rate (Mbps) &
Gain \\ \hline
{\small Benchmark 1 (No-user-cooperation)} & 3.85 & N/A & 6.58 & N/A \\
{\small Benchmark 3 (Ideal - 1 relay node)} & 3.17 & -17.7\% & 7.10 & 7.9\%
\\
{\small Benchmark 3 (Ideal - 4 relay nodes)} & 8.38 & 117.7\% & 13.70 &
108.2\% \\
{\small Proposed (4 relay nodes)} & 6.25 & 62.3\% & 10.78 & 63.9\% \\
{\small Benchmark 3 (Ideal - 9 relay nodes)} & 14.60 & 279.2\% & 20.58 &
212.8\% \\
{\small Proposed (9 relay nodes)} & 9.69 & 151.7\% & 14.54 & 121.0\% \\
\hline
\end{tabular}%
\vspace{-5mm}
\end{table*}

\subsection{How much performance gain can be achieved considering multiple
antennas at end-users?}

In order to examine the potential of our proposed framework under practical system settings, we compare Benchmark Scheme 1 and Benchmark
scheme 2 to check how much rate improvement can be achieved under infinite
D2D link capacity.

Before investigating the rate improvement by multiple antennas at end-users, we first have to examine the effects of the number of scheduled users $S$ on the rate performance, and eliminate the influence of user scheduling. Fig. \ref{fig:19cells4Tx_M} shows the cumulative
distributions of the long-term average user rates, with various number of
receive antennas $M$ and number of scheduled users $S$ when the number of
transmit antennas $L=4$. Fig. \ref{fig:19cells10Tx_M} shows the cumulative
distributions with the same parameters except for $L=10$. We notice that the
case $M=1$ stands for the Benchmark Scheme 1 without user cooperations.
Comparing Fig. \ref{fig:19cells4Tx_1} - Fig. \ref{fig:19cells4Tx_10}, we
observe that when the number of receive antennas $M$ is limited, the
relationship of rate performance and the number of scheduled users $S$ is
not straightforward. For example, as shown in Fig. \ref{fig:19cells4Tx_1},
when $L=4$ and there is no relay node, scheduling 2 users at a time ($S=2$)
achieves the best performance. While with one relay node at each user, Fig. %
\ref{fig:19cells4Tx_2} shows that the case of $S=4$ achieves the best
performance.

However, it is interesting to note that when the number of receive antennas $%
M$ exceeds a certain threshold,\footnote{Note that $M = 10$ is a reasonable setup under the 5G vision where one user per $m^{2}$ is expected.} fully loading the BS, i.e., setting the
number of scheduled users $S$ to be equal to the number of BS transmit
antennas $L$, is always the optimal strategy. For instance, we see that full
loading achieves the best SE when $M\geq 2$ in Fig. \ref{fig:19cells4Tx_M}
and when $M\geq 5$ in Fig. \ref{fig:19cells10Tx_M}. This is a practical
insight for designing user scheduling of future multiple antenna systems.

Next, we eliminate the influence of the user scheduling and focus on
examining the benefits of user cooperations. That is, for given numbers of
transmit and receive antennas $L$, $M$, we pick up the cases with the best
rate performance and then compare their SE in Fig. \ref{fig:19cells4Tx} and
Fig. \ref{fig:19cells10Tx}, respectively. We see that both figures show
great improvement in terms of rate performance of the cellular users as the
number of receive antennas $M$ increases. For example, in Fig. \ref%
{fig:19cells10Tx}, when $L=10$, an improvement up to 4.5x is observed with $%
M=10$ for the 50th percentile users compared with the case of $M=1$.

We list the rate improvement for the 10th and 50th percentile users when $%
L=4 $ in Table \ref{table:2}. A huge rate improvement is observed for both
types of users and the rate improvement increases as the number of receive
antennas $M$ increases. For instance, when $L=4$, compared with the
non-cooperative Benchmark Scheme 1, for the 10th percentile user, we obtain
a 117.7\% rate improvement when $M=5$, while 279.2\% more rate is achieved
when $M=10$. Due to the different user path loss, the rate improvement is not
linear with the number of receive antennas $M$ as indicated by MIMO theory.
However, the improvement brought by this D2D cellular framework is still
significant and useful for encountering inter-cell interference. In
addition, we also note that this improvement for low-rate users is
especially meaningful since it largely enhances the performance of cell-edge
users\ and thus improves user experience.

\begin{figure}[h]
\centering
\captionsetup[sub]{labelfont=bf,textfont=5 }
\includegraphics[width=0.95\columnwidth]{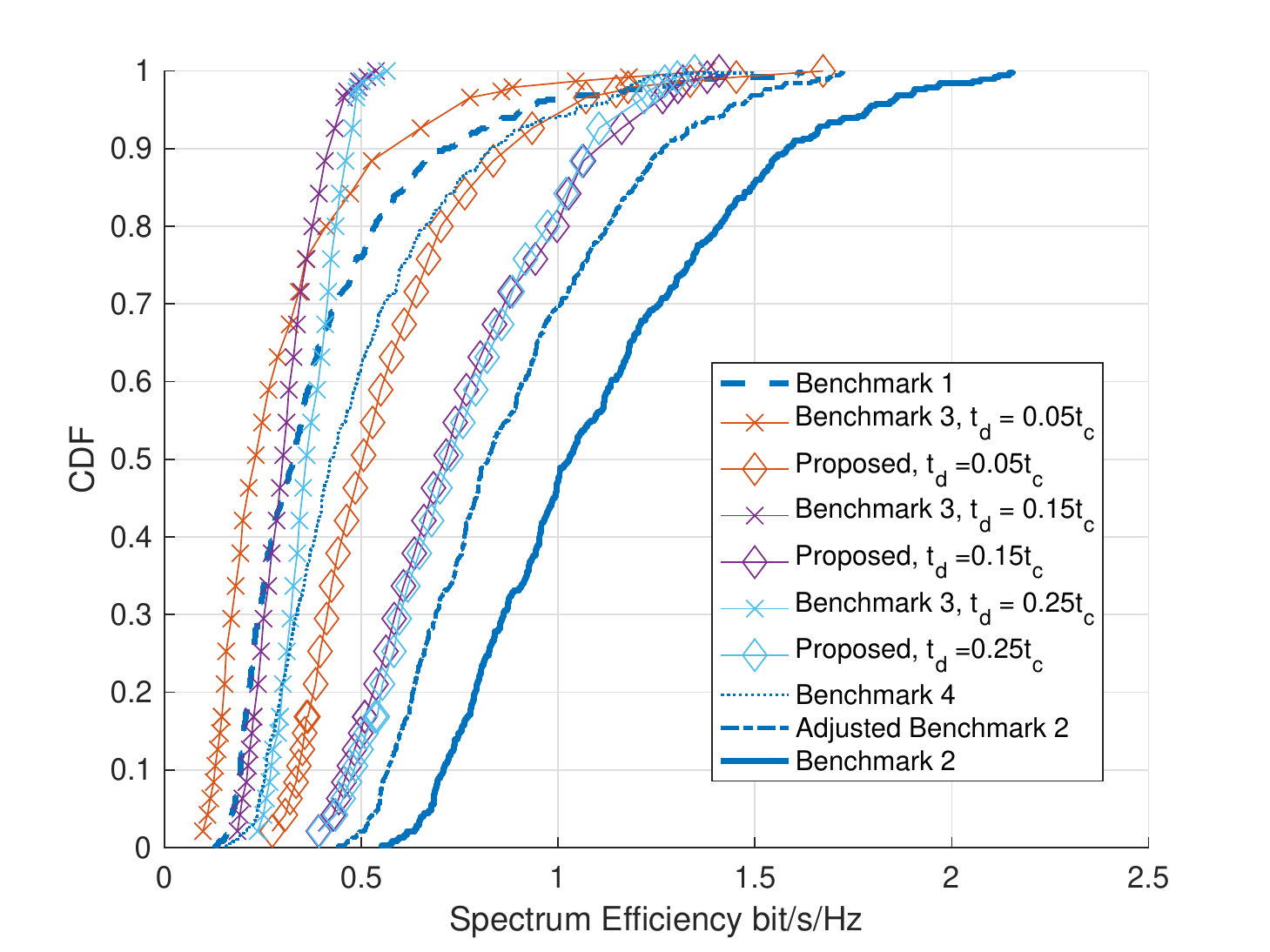}
\caption{Cumulative distribution function of user data rate comparison with
Benchmark 3, $L=4$, $M=10$.}
\label{fig:19cells4Tx9R_D2D}
\vspace{-5mm}
\end{figure}

\begin{figure}[h]
\centering
\captionsetup[sub]{font=it,labelfont={bf,it}} \includegraphics[width=0.95\columnwidth]{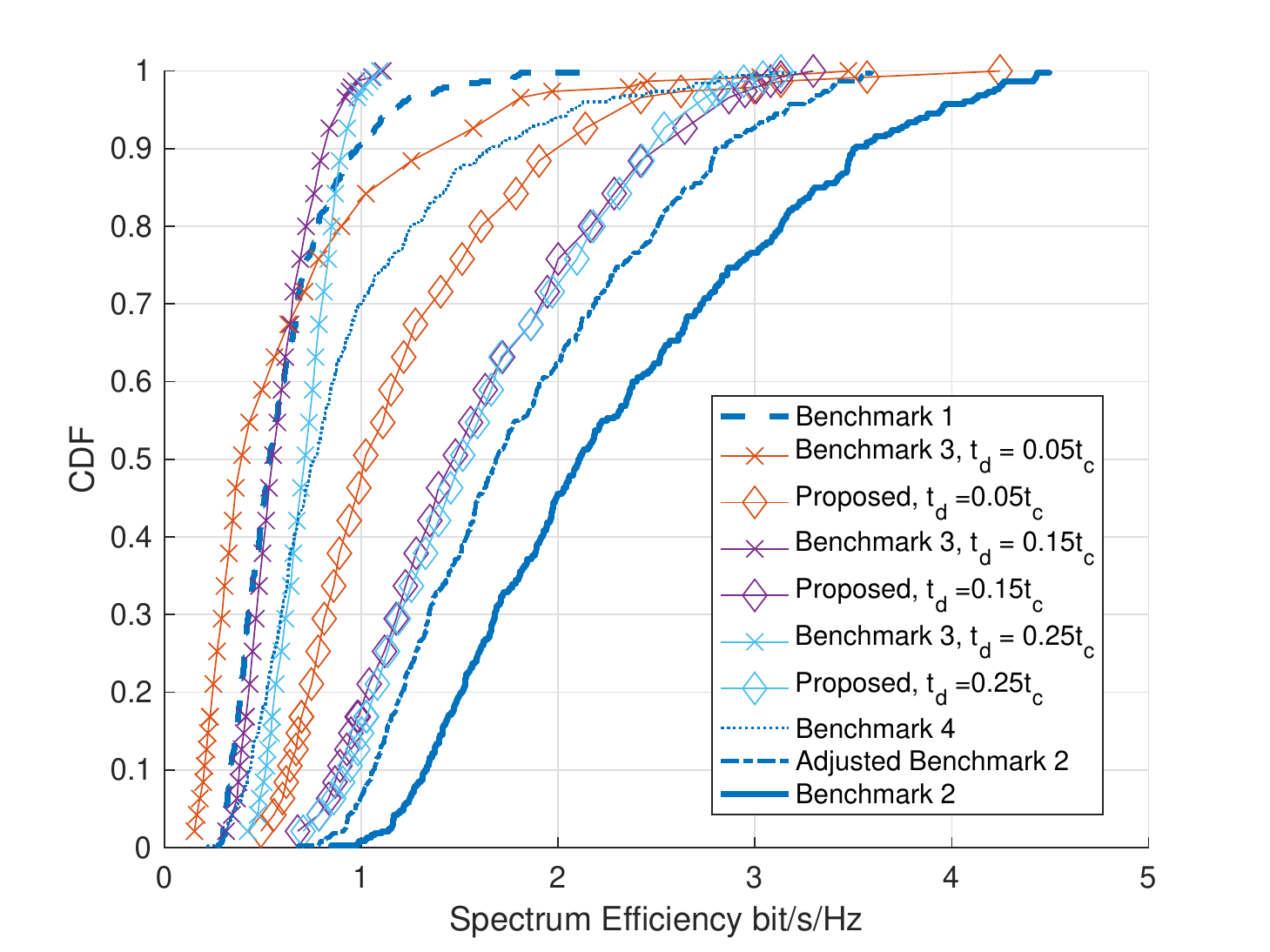}
\caption{Cumulative distribution function of user data rate comparison with
Benchmark 3, $L=10$, $M=10$.}
\label{fig:19cells10Tx9R_D2D}
\end{figure}

\subsection{The effectiveness of our proposed resource allocation}

Next, we consider the physical limits of D2D links and examine the
effectiveness of our proposed resource allocation algorithm. Fig. \ref%
{fig:19cells4Tx9R_D2D} and Fig. \ref{fig:19cells10Tx9R_D2D} compare the
cumulative distributions of the long-term average user rates for the
Benchmark Scheme 3 (equal allocation), the Benchmark Scheme 4 (multi-hop)
and our proposed algorithm. We also show the curve of Benchmark 1 as a
baseline and that of Benchmark 2 as an upper bound in the figures.

Both figures demonstrate an impressive enhancement brought by our proposed
allocation algorithm. For example, if using the equal allocation strategy,
the rate performance shows a rather limited improvement. What's worse, as
shown in Fig. \ref{fig:19cells4Tx9R_D2D}, when the D2D transmission time $%
t_{d}=0.05t_{c}$ ($t_{c}$ is the cellular frame duration), the rate might be
even degraded since the small rate increase brought by the D2D links cannot
compensate the consumed time duration for D2D links. On the other hand, we
can see that our optimized allocation guarantees an improvement of user rate
under different D2D transmission time $t_{d}$. We also observe that our
proposed algorithm, which utilizes a diversity gain, largely outperforms the
Benchmark Scheme 4 (multi-hop D2D cooperation) which achieves a power gain
under various system setups.

We also list the improvement for the 10th and 50th percentile users when $%
L=4 $ in Table \ref{table:2}. Similarly, we see that a huge rate improvement
is observed for both types of users and the rate improvement increases as
the number of relay nodes $M-1$ increases. Especially, a 2.5x improvement in
terms of the 10th percentile user rate is observed when each cellular user
is helped by 9 standby users. It is interesting to observe that the
optimized D2D allocation for $t_{d}=0.15t_{c}$ and $t_{d}=0.25t_{c}$
achieves almost the same rate performance. This can be interpreted as that
the rate increase brought by better D2D connections from $t_{d}=0.15t_{c}$
and $t_{d}=0.25t_{c}$ is cancelled by the increasing consumed time duration
on the D2D links.

We further notice that although there seems to be a big gap between the rate
performance of our optimized allocation and that of Benchmark 2, our curve is
already close to the ideal case where the D2D link capacity
is assumed to be infinite. This is because the gap here mainly comes from the average over
the total transmission time $\left( t_{d}+t_{c}\right) $ when calculating
the rate, while in Benchmark 2 we have $t_{d}=0$. If we also multiply the
ratio $\frac{t_{c}}{t_{d}+t_{c}}$ to the rate of Benchmark 2 (shown as
\textquotedblleft Adjusted Benchmark 2\textquotedblright\ in Fig. \ref%
{fig:19cells4Tx9R_D2D} and Fig. \ref{fig:19cells10Tx9R_D2D}), we can see
that the rate of our proposed method $t_{d}=0.25t_{c}$ approaches that of
Benchmark 2.

We would like to point out that the CSI from the BS to relay nodes is
essential in our scheme. This is because without the relay CSI, the transmit
beamformer is still designed based on a single receive antenna model, thus the interference cannot be effectively managed. Numerical results also confirm this and we do not show here due to the length limit.

\section{Conclusion}

The integration of local device-to-device (D2D) communications and cellular
connections has been intensively studied to meet the co-existing requirement
of D2D and cellular communications. In this paper, we exploited the D2D
communication capability of standby users and proposed a hybrid D2D-cellular
scheme applying the distributed MIMO technology to improve the data rate and
user experience for cellular users. More specifically, a virtual antenna
array was formed by sharing antennas across different terminals to realize
the diversity gain of MIMO systems. To achieve this, mmWave D2D links were
considered to enable high data rate D2D communications. We then designed a
D2D multiple access protocol and formulated the optimization problem of
joint cellular and D2D resource allocation for downlink transmissions of our
proposed scheme. We obtained a closed-form solution for the D2D resource
allocation, which revealed useful insights for D2D resource allocation.
Extensive system-level simulations were performed to demonstrate the
effectiveness of the proposed scheme.

\end{document}